\documentstyle{amsppt}
\mag=\magstep1
\pagewidth{160truemm} \pageheight{235truemm}
\NoBlackBoxes
\rightheadtext{Quantum Mechanics and Relativity}
\TagsOnRight

\define\F{\flushpar}

\define\q{\quad}

\define\MP{\medpagebreak}
\define\BP{\bigpagebreak}

\define\R{{\Bbb R}}
\define\N{{\Bbb N}}
\define\HH{{\Cal H}}
\define\UU{{\Cal U}}
\define\OO{{\Cal O}}
\define\la{{\lambda}}

\define\Ltnn{{L^2(\R^{3n})}}

\define\tT{{\tilde T}}
\define\tH{{\tilde H}}

\define\SS{{\Cal S}}

\rightline{KIMS-1996-12-08}
\rightline{gr-qc/9612043}
\BP

\vskip10pt

\topmatter
\title    Quantum Mechanics and Relativity\\
--- Their Unification by Local Time ---
\endtitle
\author Hitoshi Kitada
\endauthor
\affil
Department of Mathematical Sciences\\
University of Tokyo\\
Komaba, Meguro, Tokyo 153, Japan\\
e-mail: kitada\@ms.u-tokyo.ac.jp
\endaffil
\abstract
 In a framework of a stationary universe, time is defined as a local
 and quantum-mechanical notion in the sense that it is defined for
 each local and quantum-mechanical system consisting of a finite
 number of particles. In this context, the total universe consisting
 of an infinite number of particles has no time associated, and quantum
 mechanics and general theory of relativity are united consistently.
 Relativistic Hamiltonians including gravitation are derived as a
 consequence of our treatment of observation. Related open problems
 in mathematical physics are presented.
\endabstract
\endtopmatter

\document


\subhead\nofrills{Introduction}
\endsubhead

\vskip 12pt

\F
Physics is a work to explain phenomena, i.e. a job to give a
 description of visible events. Insofar as we understand physics as
 such activities, it is neither surprise nor ridiculous thing if one
 takes other ways in explaining phenomena than the present physics:
The problem of combining relativity and quantum theories, which has
 been an old and difficult one, might be able to be considered from
 a different viewpoint than the present trends where relativity
 theory is tried to be quantized or quantum mechanics is tried to be
 modified relativistically. It is enough if one can explain
 relativistic quantum-mechanical phenomena or
 observations of them, in a systematic way.
 The purpose of the present paper is to give an attempt in this
 direction to explain relativistic quantum-mechanical phenomena. To
 make clear the contrast of our approach to the current physics, we
 briefly review the problems of physics in relation with
 relativity and quantum theories.

As is well-known, the solution $\psi=\psi(t)=\psi(x,t)$ $(x\in\R^{3N})$
of the Schr\"odinger equation for $N(\ge 1)$ particles, numbered as
 $\ell=1,2,\cdots,N$,
with positions $x_\ell=(x_{\ell1},x_{\ell2},x_{\ell3})\in \R^3_{x_\ell}$
 and masses $m_\ell>0$:
$$  
\aligned
\frac{1}{i}\frac{d \psi}{d t}(t)&+H\psi(t)=0,\quad
\psi(0)=\phi,\q \phi\in{\Cal D}(H)\subset L^2(\R^{3N}),  \\ 
H&=-\sum_{\ell=1}^N\sum_{k=1}^3 \frac{1}{2m_\ell}
\frac{\partial^2}{\partial x_{\ell k}^2}
+\sum_{1\le i<j\le N}V_{ij}(x_i-x_j),
\endaligned
\tag S
$$
is invariant with respect to the Galilei transformation:
$$
\split
&x^\prime_{\ell}=x_{\ell}-v t,\q (\ell=1,2,\cdots,N)\\
&t'=t
\endsplit
$$
up to a factor of absolute value 1:
$$
\exp\left[i\sum_{\ell=1}^N \left(\frac{1}{2}m_\ell v^2 t 
-m_\ell v\cdot x_{\ell}\right)\right],
$$
 consistently with Born's interpretation ([Bo]).
Here we adopted the unit system such that $\hbar=\frac{h}{2\pi}=1$;
 $v=(v_1,v_2,v_3)\in\R^3$ denotes the velocity between two
 inertial frames of reference; and 
$v\cdot x_\ell=\sum_{k=1}^3 v_k x_{\ell k}$ is the inner
 product of $v$ and $x_\ell$. This implies that the Schr\"odinger
 equation is not invariant under Lorentz transformation: 
$x^{\mu\prime}=a^\mu_{\ \nu} x^\nu$
from $\R_t\times\R^3_{x_\ell}$ to $\R_{t'}\times\R^3_{x'_\ell}$
with $x^{0\prime}=ct'$, $x^{k\prime}=x'_{\ell k}$, $x^0=ct$, and
$x^k=x_{\ell k}$. Here $c$ is the speed of light in vacuum
 and the coefficients $a^\mu_{\ \nu}$ are independent
 of $\ell$ and determined by the following condition with some extra
 informations:
$$
(x^{1\prime})^2+(x^{2\prime})^2+(x^{3\prime})^2-(x^{0\prime})^2
=(x^1)^2+(x^2)^2+(x^3)^2-(x^0)^2.
$$
 Therefore the quantum mechanics which is described by Schr\"odinger
 equation is understood, in the current physical context, as
 incompatible with special theory of relativity. Even if
 the free energy part in (S) (i.e. the sum of
 the differential operators in (S)) is replaced by a Lorentz
 invariant one as we will do in (QMG) of subsection I.3.2,
 the Schr\"odinger equation is not Lorentz invariant.
 This fact is known as that the instantaneous force
 among particles, which depends only on the locations of those particles,
 is not relativistic, i.e. is not Lorentz invariant.

One of the features of the Schr\"odinger equation is that it yields
 the stability of matter (of the first kind, see [Li]), which is
 violated in the classical framework of Maxwell's equations and
 Rutherford model of atoms, where atoms collapse by the continuous
 radiation of light from the electrons around the nucleus according
 to classical electromagnetism so that the electrons fall
 into the nucleus. However the Schr\"odinger operator $H$ defined in
 (S) is bounded from below by some constant $- L > - \infty$ in the
 sense that $L^2$-inner product $(Hf,f)$ satisfies
$$
(Hf,f) \ge -L (f,f)
$$
for any $f$ belonging to the domain ${\Cal D}(H)$ of $H$
under a suitable assumption on the pair potentials $V_{ij}$. 
This means that the total energy of the quantum system does not
 decrease below $- L$, therefore the system does not collapse. In
 this respect, quantum mechanics remedies the difficulty of
 classical theory, while it is not Lorentz invariant.

In 1928, Dirac [Di] introduced a system of equations, which is
 invariant under Lorentz transformation, and could explain some of
 the relativistic quantum-mechanical phenomena. However, Dirac
 operator is not bounded from below, and Dirac equation does not
 imply the stability of matter unlike the Schr\"odinger equation.

Dirac thus proposed an idea that the vacuum is filled with electrons
 with negative energy so that the electrons around the nucleus
 cannot fall into the negative energy anymore by the Pauli
 exclusion principle, which explains the stability of matter.
 However, if one has to consider plural kinds of elementary
 particles at a time, one has to introduce the vacuum which
 is filled with those plural kinds of particles, and the vacuum
 can depend on the number of the kinds of particles which one
 takes into account. The vacuum then may not be determined to be unique.
 Further if one has to include Bosons into consideration,
 the Pauli exclusion principle does not hold and the stability
 of matter does not follow. In this sense, the idea of
 ````filling the negative energy sea;" unfortunately,"  ``is
 ambiguous in the many-body case," as Lieb writes in
 [Li, p.33].

Quantum field theory is introduced (see [St] for a review) 
to overcome this difficulty as well as to explain the
 annihilation-creation phenomena of particles, which are
 familiar in elementary particle physics. Quantum field
 theory is a theory of infinite degrees of freedom. In
 the case of the Schr\"odinger equation (S), the
 degree of freedom is $3N$, the number of coordinates 
$x_{11}, x_{12}, x_{13}, \cdots, x_{N1}, x_{N2}$, and $x_{N3}$ of 
$N$ particles. Contrary to this, quantum field theory deals with
 the infinite number of particles, which makes it possible to
 discuss creation-annihilation processes inside the theory.
 However, since it deals with infinite number of freedom, even at
 the first step of the definition of the Hamiltonian of the system
 obtained by second quantization, there is a difficulty, the
 difficulty of divergence. This sort of difficulty appears at almost
 every stage of the development of the theory, and physicists had 
to find clever ways to avoid the difficulties at each step after the
 theory was introduced. Mathematically, the difficulty of divergence
 has not been overcome yet at all. Physicists however noticed that
 if one could get finite quantities in a systematic way by 
extracting some infinite quantities from the divergent quantities,
 then those finite quantities might express the reality. Actually 
in their explanation of Lamb shift, they seemed to have succeeded
 going in this way and to have been able to give predictions
 outstandingly close to experiments. However, the calculation
 done is up to the 6th or 8th order of a series giving Lamb shift
 or anomalous magnetic moment of electrons ([K-L]). Dyson noticed
 ([Dy]) that the series has symptom to diverge to infinity. 

The procedure mentioned in the above to yield finite quantities from
 infinite ones is called process of ``renormalization," and still
 forms active areas of researches in theoretical
 physics. In the mathematical attempt, called ``axiomatic quantum
 field theory," which was planned to clarify the meaning of 
 quantum field theory and construct the theory consistently, it is
 known that in some mathematical but important examples (see, e.g.,
 [Fr]), renormalizability conditions and the axioms of quantum
 field theory yield that the theory must not involve interaction
 terms inside the theory. I.e., the theory is void as a physics.

These are the situation currently understood as an incompatibility
 problem between quantum theory and special theory of relativity. In
 the case of general theory of relativity and quantum mechanics,
 the situation seems similar or no better (see, e.g., [Ish] for a 
review of the current approaches). The traditional attempts toward
 the unification of quantum theory and general relativity, like
 quantum gravity, superstring theory, and so on, are trying to find
 a way to unify them in a single layered theory where these two
 difficult theories should admit each other.

We present below an attempt in a different direction, where
 general theory of relativity and quantum mechanics are considered
 as independent aspects of nature, but as playing complementary
 roles to each other. Our approach may be called a two-layered
 theory, where these two theories have their own residences and they
 interfere only when observation is done. A procedure which
 describes the interference between them at observation will be our
 basis of explanation of relativistic quantum-mechanical phenomena.

Our spirit behind the procedure we will introduce below for that
 interference is that what is intrinsic is the quantum-mechanical
 aspect of nature, while relativity plays a role of glasses
 to see nature. This attitude is contrary to the one adopted
 by current physics, which in origin comes from the spirit of
 Einstein [Ein]:
\block
 Thus, according to the general theory of relativity, gravitation
 occupies an exceptional position with regard to other forces,
 particularly the electromagnetic forces, since the ten functions
 representing the gravitational field at the same time define the
 metrical properties of the space measured.
\endblock
 His position is that the metrical properties of space-time are 
 intrinsic for nature, and space-time is a vessel of nature,
 into which other forces should be incorporated. In the framework
 of classical theory, electromagnetic forces can be treated
 in this direction in the sense that the equation for 
 electromagnetic fields can be written as a tensor equation.
 In the framework of quantum theory, the characteristic of
 traditional approach is to treat gravity as a one which should
 be quantized, and the inclusion of other forces
 is a problem which is treated only after gravity is
 quantized successfully. In such attempts to quantize gravity
 or general theory of relativity, the canonical formalism of
 general theory of relativity is assumed usually, and this means
 that one has to introduce some global time coordinate which
 is common throughout the total universe. This itself produces
 a problem incompatible with the spirit of general theory of
 relativity that time is a local notion. If one would
 admit of introducing such a global time, it is difficult to
 reformulate general theory of relativity into canonical
 formalism even if gravitation is weak (see, e.g., [Ish]),
 and the quantization of gravity or space-time remains as
 a difficult problem even if one would defer to the global time.

 To overcome these difficulties, we introduce a notion of local
 time $t_L$ which is proper to each local system $L$ consisting of
 a finite number of quantum-mechanical particles. Our local time is
 a {\it quantum-mechanical} notion inasmuch as it is defined
 in each quantum-mechanical system as a parameter $t_L$ in the
 exponent of the propagator $\exp(-it_L H_L)$ describing the
 propagation of the local system $L$. It is a {\it local}
 notion defined for each local system $L$ with a local Hamiltonian
 $H_L$, and this will enable us to regard the time $t_L$ as a
 {\it classical general relativistic} local time, proper to the
 {\it center of mass} of the local system $L$, by identifying
 the classical particles with those centers of mass of local systems.
 We will show that these classical local times proper to the centers
 of mass of local systems constitute general relativistic notion
 of local times, compatible with the quantum mechanics inside each
 local system. The proof is, in part, a recall of the 
 inclusion/exclusion assumption which has been
 adopted in physics, that the time $t_B$ of a
 bigger system $L_B$ which includes a smaller system $L_S$ dominates
 the time $t_S$ of $L_S$, i.e. the assumption that $t_S$ must be
 equal to $t_B$ if the system $L_B$ includes $L_S$.
 Apart from this traditional position on which
 physics has been founded, we retrieve the independence of each
 local system and its time coordinate among local systems, and
 liberate them from the bondage of inclusion/exclusion relation,
 which has been implicitly assumed for systems of physical
 particles. Geometrically expressed, our position may be
 formulated as a vector bundle with base space $X$ representing
 the Riemannian manifold consisting of classical particles,
 identified with the centers of mass of local systems, and with
 the local system $L$ which obeys the quantum mechanics on its own
 geometry being associated as a fibre to each point $x \in X$,
 which is identified with the center of mass of the local system
 $L$.

 There is a theory by Prugove\v cki [Pru] successful, in a sense,
 in quantizing general relativity, where he modifies quantum
 mechanics and general relativity so that the usual results are obtained
 as limits of his theory. His approach looks similar to ours in that he
 associates a {\it Lorentzian} quantum-mechanical world to each point
 of a Riemannian manifold as a fibre, regarding the total universe as
 a vector or fibre bundle equipped with connections
 compatible with the Riemannian metric of the base Riemannian space.
 Our approach differs from his in the following
 points:
\roster
\item Each quantum-mechanical world associated to a point of a
 Riemannian manifold is {\it Euclidean}; 
\item We do not introduce any
 connections among those Euclidean quantum mechanics; 
\item We treat
 electromagnetic forces and gravity on the same level in our
 explanation of observation, under the assumption that
 gravitation is weak; and
\item Quantum mechanics
 and general theory of relativity are intact in our formulation.
\endroster
 The explanation of observation stated in the third item is
 our point and is realized by a procedure which yields
 a quantized Hamiltonian including gravity and electric
 forces on the same level.

\hyphenation{deform}

In our explanation of observation, we appeal to a procedure
 which transforms \linebreak quantum-mechanical quantities
 into the classical quantities which obey the relativistic change
 of coordinates among the Euclidean quantum-mechanical worlds,
 which we call local systems. The quantum-mechanical world associated
 to each point of a Riemannian manifold has no relation with the
 Riemannian metric of the base space of our vector bundle, for we
 do not define any connections among the quantum-mechanical worlds.
The procedure which transforms 
 quantum-mechanical quantities into classical quantities is consistent
with the two aspects of nature, i.e. with the quantum-mechanical
 aspect inside local systems and the general relativistic aspect
 outside local systems, because the results obtained by transformations
 are just concerned with observed facts. The relativity
 appears in our theory as ``glasses," which deform
 quantum-mechanical quantities into classical relativistic quantities
at each step of quantum-mechanical evolution, so that
the resultant classical quantities accord with the observation.
 The intrinsic for our theory is quantum mechanics inside
 local systems, and relativity modifies quantum-mechanical calculations
 to accord with observations.

\hyphenation{ex-clu-sion}

 Summing up, our point is in the liberation of local systems from
 the inclusion/exclu-sion relation which has been an implicit assumption
 of physics. Instead of the inclusion/exclusion relation, we introduce
 a relation which transforms the quantum-mechani-cal values
 to classical relativistic values, as a procedure describing
 the interference between the two aspects of nature, the general
 relativistic aspect and the quantum-mechanical aspect.

In Part I of the paper, we give a presentation of our theory without
 using the notion of vector bundle. We first recall in section I.1
 the basic notions related with the
 definition of local times from [Ki]. This notion of local times is
 a quantum-mechanical one defined in each local system consisting of
 a finite number of quantum-mechanical particles. In the sense that
 the local time is a local notion, it serves an ingredient which
 adheres the two layers: general theory of relativity and
 quantum mechanics. A result in many body quantum scattering is
 used to assign the usual meaning of time to our notion of local
 times. In section I.2, we review the proof of the consistency of the
 notion of local times with general theory of relativity. We give,
 in section I.3, a procedure of interpreting observation of
 quantum-mechanical process through the glasses of the relativity,
 yielding a relativistic quantum-mechanical Hamiltonian which explains
 gravitation and electric forces in quantum-mechanical way.
In Part II, we treat two examples following the spirit of Part I.
We present some open problems related with our formulation of physics
in Part III.

\BP

\vskip16pt

\centerline{\bf Part I. Local Time and Observation}

\BP

\vskip12pt

\subhead\nofrills{I.1. Local Time}
\endsubhead

\BP

\F
 To state our definition of quantum-mechanical local times, we begin
 with introducing a stationary universe $\phi$. What we adopt here
 for the universe may be called a closed universe, within which
 is all and which has a definite property specified by a certain
 quantum-mechanical condition.

Let $\HH$ be a separable Hilbert space, and set
$$
\UU=\{\phi\}=\bigoplus_{n=0}^\infty \left(\bigoplus_{\ell=0}^\infty
\HH^n  \right) \q (\HH^n=\underbrace{\HH\otimes\cdots\otimes
\HH}_{n\ \text{factors}}).
$$
$\UU$ is called a Hilbert space of possible universes. An element 
$\phi$ of $\UU$ is called a universe and is of the form of an
 infinite matrix $(\phi_{n\ell})$ with components $\phi_{n\ell} 
\in \HH^n$. $\phi=0$ means $\phi_{n\ell}=0$ for all $n, \ell$.

Let $\OO=\{ A\}$ be the totality of the selfadjoint operators $A$
 in $\UU$ of the form $A\phi=(A_{n\ell}\phi_{n\ell})$ for
$\phi=(\phi_{n\ell})\in{\Cal D}(A)\subset\UU$, where each component
 $A_{n\ell}$ is a selfadjoint operator in $\HH^n$. We assume the
 following condition for our universe $\phi$.

{\bf Axiom 1.}\ There is a selfadjoint operator $H\in\OO$ in 
$\UU$ such that for some $\phi\in \UU-\{0\}$ and $\la\in \R$
$$
H\phi=\la \phi \tag U
$$
in the following sense: Let $F_n$ be a finite subset of 
${\N}=\{1,2,\cdots\}$ with $\sharp(F_n)(=$ the number of elements in
 $F_n)=n$ and let $\{ F_n^\ell\}_{\ell=0}^\infty$ be a countable set of
 such $F_n$. Then the formula (U) in the above means that there are
 integral sequences $\{n_k\}_{k=1}^\infty$ and 
$\{ \ell_k\}_{k=1}^\infty$ and a real sequence 
$\{\la_{n_k \ell_k}\}_{k=1}^\infty$
such that 
$F_{n_k}^{\ell_k}\subset F_{n_{k+1}}^{\ell_{k+1}}$; 
$\bigcup_{k=1}^\infty F_{n_k}^{\ell_k} = \N$;
$$
H_{n_k\ell_k}\phi_{n_k\ell_k}=\la_{n_k\ell_k}\phi_{n_k\ell_k},
\q \phi_{n_k\ell_k}\ne0,\q k=1,2,3,\cdots; \tag Eigen
$$
and
$$
\la_{n_k\ell_k}\to\la\q \text{as}\q k\to\infty.
$$

 $H$ is an infinite matrix $(H_{n\ell})$ of selfadjoint operators
 $H_{n\ell}$ in $\HH^n$. Axiom 1 asserts that this matrix converges
 in the sense of (U) on our universe $\phi$. We remark that our
 universe $\phi$ is not determined uniquely by this condition.

 The universe as a state $\phi$ is a whole, within which is all.
 As such a whole, the state $\phi$ can follow the two ways: The one
 is that $\phi$ develops along a global time $T$ in the grand
 universe $\UU$ under a propagation $\exp(-iTH)$, and another is
 that $\phi$ is a bound state of $H$. If there were such a global
 time $T$ as in the first case, all phenomena had to develop along
 that global time $T$, and the locality of time would be lost. We
 could then {\it not} construct a notion of local times compatible
 with general theory of relativity. The only one possibility is
 therefore to adopt the stationary universe $\phi$ of Axiom 1.

\BP

The following axiom asserts the existence of configuration and
 momentum operators and that the canonical commutation relation
 between them holds. This is a basis of our definition of time,
 where configuration and momentum are given first, and then local
 times are defined in each local system of finite number of 
quantum-mechanical particles.

\MP

{\bf Axiom 2.}\ Let $n\ge 1$ and $F_{n+1}$ be a finite subset of 
${\N}=\{1,2,\cdots\}$ with $\sharp(F_{n+1})=n+1$. Then for any 
$j\in F_{n+1}$, there are selfadjoint operators 
$X_j =(X_{j 1},X_{j2},X_{j3})$ and $P_j =(P_{j1},P_{j2},P_{j3})$ in
$\HH^n$, and constants $m_j>0$ such that
$$
[X_{j\ell},X_{k m}]=0,\q [P_{j\ell},P_{k m}]=0,
\q [X_{j\ell},P_{k m}]=i\delta_{jk}\delta_{\ell m},
$$
$$
\sum_{j\in F_{n+1}} m_j X_j=0,\q \sum_{j\in F_{n+1}} P_j=0.
$$

The Stone-von Neumann theorem and Axiom 2 specify the space
 dimension (see  [A-M], p.452) as 3 dimension. We identify $\HH^n$ 
with $\Ltnn$ in the following.

\BP

What we want to mean by the $(n,\ell)$-th component $H_{n\ell}$ 
$(n,\ell\ge 0)$ of $H=(H_{n\ell})$ in Axiom 1 is the usual $N=n+1$
 body Hamiltonian with center of mass removed in accordance with the
 requirement $\sum_{j\in F_{n+1}} m_j X_j=0$ in Axiom 2. For the
 local Hamiltonian $H_{n\ell}$ we thus make the following postulate.

\MP

{\bf Axiom 3.}\ The component Hamiltonian $H_{n\ell}$ $(\ell\ge0)$
 of $H$ in Axiom 1 is of the form
$$
H_{n\ell}=H_{n\ell0}+V_{n\ell},\q V_{n\ell}=\sum\Sb\alpha=(i,j)\\
 1\le i<j<\infty, i,j\in F_N^\ell\endSb V_\alpha(x_\alpha)
$$
on $C_0^\infty(\R^{3n})$, where $x_\alpha=x_i-x_j$ with $x_i$ being
 the position vector of the $i$-th particle, and 
$V_\alpha(x_\alpha)$ is a real-valued measurable function of 
$x_\alpha\in \R^3$ which is $H_{n\ell0}$-bounded with 
$H_{n\ell0}$-bound of $V_{n\ell}$ less than 1.
$H_{n\ell0}=H_{(N-1)\ell0}$ is the free Hamiltonian of the 
$N$-particle system, whose concrete form is similar to the
 interaction-free part of $H$ in (S) of the introduction.

This axiom implies that $H_{n\ell}=H_{(N-1)\ell}$ is uniquely
 extended to a selfadjoint operator bounded from below in 
$\HH^n=\HH^{N-1} =L^2(\R^{3(N-1)})$ by the Kato-Rellich theorem.

We do not include vector potentials in the Hamiltonian 
$H_{n\ell}$ of Axiom 3, for we take the position that what 
is elementary is the electronic charge, and the magnetic 
forces are the consequence of the motions of charges. Thus 
when we restrict our attention to a system consisting of 
the $N$ number of particles, the vector potential is 
redundant to our argument. It would be, however, a good
 approximation to introduce vector potentials, when we 
consider a subsystem of a bigger system, and we concentrate 
on the analysis of the behavior of that subsystem inside the
 bigger system.

\BP

Let $P_H$ denote the orthogonal projection onto the space of
bound states for a selfadjoint operator $H$. We recall that a state
 orthogonal to the space of bound states is called a scattering
 state. Let $\phi=(\phi_{n\ell})$ with 
$\phi_{n\ell}=\phi_{n\ell} (x_1,\cdots,x_n)\in \Ltnn$ 
be the universe in Axiom 1, and let $\{n_k\}$ and $\{\ell_k\}$ be
 the sequences specified there. Let $x^{(n,\ell)}$ denote the
 relative coordinates of $n+1$ particles in $F_{n+1}^\ell$.
\MP

\F
{\bf Definition 1.}
\roster
\item
We define $\HH_{n\ell}$ as the sub-Hilbert space of $\HH^n$
 generated by the functions $\phi_{n_k\ell_k}$ $(x^{(n,\ell)},y)$ of
 $x^{(n,\ell)}\in \R^{3n}$ with regarding $y\in\R^{3(n_k-n)}$ as a
 parameter, where $k$ moves over a set $\{k\ |\ n_k\ge n, 
F_{n+1}^\ell\subset F_{n_k+1}^{\ell_k}, k\in \N\}$.
\item
$\HH_{n\ell}$ is called a {\it local universe} of  $\phi$.
\item
$\HH_{n\ell}$ is said to be non-trivial if 
$(I-P_{H_{n\ell}})\HH_{n\ell}\ne\{0\}$.
\endroster

The total universe $\phi$ is a single element in  $\UU$. The local
 universe $\HH_{n\ell}$ may be richer and may have elements more
 than one. This is because we consider the subsystems of the
universe consisting of a finite number of particles. These
 subsystems receive the influence from the other particles of
 infinite number outside the subsystems, and may vary to constitute
 a non-trivial subspace $\HH_{n\ell}$. 

\hyphenation{Coulomb}

To state this mathematically, let us assume that the pair potentials 
are of the Coulomb type $V_{ij}(x_{ij})=c_{ij}|x_{ij}|^{-1}$ 
$(c_{ij}\in \R)$, which are the typical examples we had in mind in 
Axiom 3. Consider, e.g., a system $H_{3\ell}$ consisting of four 
particles, the one of which has positive charge, and other three 
have negative charge. Then this system tends to scatter, i.e. 
it is probable that this system is in a scattering state 
with respect to the Hamiltonian  $H_{3\ell}$ (see, e.g., [Cy, p.50] 
for a theorem asserting the absence of eigenvalues for a similar case).
Add one particle with positive charge to this system $H_{3\ell}$ to 
constitute a system $H_{n_k\ell_k}$ $(n_k=4)$. Then this new system 
may be in an eigenstate $\phi_{n_k\ell_k}=\phi_{4\ell_k}$ 
with respect to the extended Hamiltonian $H_{n_k\ell_k}=H_{4\ell_k}$ 
for {\it some} eigenvalue $\lambda_{n_k\ell_k}=\lambda_{4\ell_k}$ so that
 it satisfies the condition (Eigen) in the above for a $k$,
 while the restriction $\phi_{4\ell_k}(x^{(3,\ell)},y)$ to
 $\R^{9}_{x^{(3,\ell)}}$, with $y\in\R^3$ arbitrary but fixed,
 of the bound state
 $\phi_{4\ell_k}$ of $H_{4\ell_k}$ is a scattering state of
 the original system $H_{3\ell}$. Here $y\in\R^3$ is the
 intercluster coordinates between the added particle of
 positive charge and the center of mass of the four
 particles in the system $H_{3\ell}$. Namely, the extended system
 $H_{4\ell_k}$ is in a bound state $\phi_{4\ell_k}$, while the
 restriction $\phi_{4\ell_k}(x^{(3,\ell)},y)$ moves over
 the scattering states of $H_{3\ell}$ belonging to the
 Hilbert space $L^2(\R^{9}_{x^{(3,\ell)}})$ of the state
 vectors for the system $H_{3\ell}$, and constitutes a
 nontrivial subspace $\HH_{3\ell}$ of $\HH^3$ when $y$ varies.

\BP

\F
{\bf Definition 2.}
\roster
\item
The restriction of $H$ to $\HH_{n\ell}$ is also denoted by the same
 notation  $H_{n\ell}$ as the $(n,\ell)$-th component of $H$.
\item
We call the pair
$(H_{n\ell},\HH_{n\ell})$  a local system.
\item
The unitary group
$e^{-itH_{n\ell}}$ $(t\in \R^1)$ on $\HH_{n\ell}$ is called the 
{\it proper clock} of the local system $(H_{n\ell},\HH_{n\ell})$, if
 $\HH_{n\ell}$ is non-trivial: 
$(I-P_{H_{n\ell}})\HH_{n\ell}\ne \{0\}$.
(Note that the clock is defined only for $N=n+1\ge 2$, since 
$H_{0\ell}=0$ and $P_{H_{0\ell}}=I$.)
\item
The universe $\phi$ is called {\it rich} if $\HH_{n\ell}$ equals
$\HH^n=L^2(\R^{3n})$ for all $n\ge 1$, $\ell\ge0$. For a rich
 universe $\phi$, $H_{n\ell}$ equals the $(n,\ell)$-th component of
 $H$.
\endroster

\F
{\bf Definition 3.}
\roster
\item
The parameter $t$ in the exponent of the proper clock 
$e^{-itH_{n\ell}}=e^{-itH_{(N-1)\ell}}$ of a local system 
$(H_{n\ell},\HH_{n\ell})$ is called the (quantum-mechanical) 
{\it proper time} or {\it local time} of the local system 
$(H_{n\ell}, \HH_{n\ell})$, if $(I-P_{H_{n\ell}})\HH_{n\ell}\ne
 \{0\}$. 
\item
This time $t$ is denoted by $t_{(H_{n\ell},\HH_{n\ell})}$ indicating
 the local system under consideration.
\endroster

This definition is a one reverse to the usual definition of the
 motion or dynamics of the $N$-body quantum systems, where the time
 $t$ is given {\it a priori} and then the motion of the particles is
 defined by $e^{-itH_{(N-1)\ell}}f$ for a given initial state $f$ of
 the system.

{\it Time} is thus defined only for local systems
$(H_{n\ell},\HH_{n\ell})$  and is determined by the associated
 proper clock $e^{-itH_{n\ell}}$. Therefore there are infinitely
 many number of times $t=t_{(H_{n\ell},\HH_{n\ell})}$ each of which
 is proper to the local system $(H_{n\ell},\HH_{n\ell})$. In this
 sense time is a local notion. There is no time for the total
 universe $\phi$ in Axiom 1, which is a bound state of the total
Hamiltonian $H$ in the sense specified by the condition (U) of 
\linebreak Axiom 1.

\BP

 To see the meaning of our definition of time, we quote a theorem
 from [En]. To state the theorem we make some notational 
 preparation concerning the local system 
$(H_{n\ell},\HH_{n\ell})$, assuming that the universe $\phi$ is 
rich: Let $b=(C_1,\cdots,C_{\sharp(b)})$ be a decomposition
 of the set $\{1,2,\cdots,N\}$ $(N=n+1)$ into $\sharp(b)$ number of
 disjoint subsets $C_1,\cdots,C_{\sharp(b)}$ of $\{1,2,\cdots,N\}$.
 $b$ is called a cluster decomposition. 
 $H_b=H_{n\ell,b}=H_{n\ell}-I_b=H^b_{n\ell}+T_{n\ell,b}=H^b+T_b$ is
 the truncated Hamiltonian for the cluster decomposition $b$ with 
 $1\le\sharp(b)\le N$, where $I_b$ is the sum of the intercluster
 interactions between various two different clusters in $b$, and 
 $T_b$ is the sum of the intercluster free energies
 among various clusters in $b$.  $x_b\in {\Bbb R}^{3(\sharp(b)-1)}$ 
is the intercluster coordinates among the centers of mass of the
 clusters in $b$, while $x^b\in {\Bbb R}^{3(N-\sharp(b))}$ denotes the
 intracluster coordinates inside the clusters of $b$ so that 
$x\in {\Bbb R}^{3n}={\Bbb R}^{3(N-1)}$ is expressed as $x=(x_b,x^b)$.
 Note that $x^b$ is decomposed as $x^b=(x^b_1,\cdots,x^b_{\sharp(b)})$,
 where each $x^b_j\in {\Bbb R}^{3(\sharp(C_j)-1)}$ is the internal
 coordinate of the cluster $C_j$, describing the configuration of
 the particles inside $C_j$. The operator $H^b$ is accordingly
 decomposed as $H^b=H_1+\cdots+H_{\sharp(b)}$, and each component
 $H_j$ is defined in the space 
$\HH^b_j=L^2({\Bbb R}^{3(\sharp(C_j)-1)}_{x^b_j})$, whose tensor
 product $\HH^b_1\otimes\cdots\otimes\HH^b_{\sharp(b)}$ is the
 internal state space $\HH^b=L^2({\Bbb R}^{3(N-\sharp(b))}_{x^b})$.
 The free energy $T_b$ is defined in the external space 
$\HH_b=L^2({\Bbb R}^{3(\sharp(b)-1)}_{x_b})$, and the truncated
 Hamiltonian $H_b=H^b+T_b=I\otimes H^b+T_b\otimes I$ is defined
 in the total space 
$\HH_{n\ell}=\HH_b\otimes \HH^b=L^2({\Bbb R}^{3(N-1)}_x)$.
$v_b$ is the velocity operator conjugate to the
 intercluster coordinates $x_b$.
 $P_b=P_{H^b}$ is the eigenprojection associated with the subsystem
 $H^b$ of $H$, i.e. the orthogonal projection onto the eigenspace of
 $H^b$, defined in $\HH^b$ and extended obviously to the total space
 $\HH_{n\ell}$. $P_b^M$ is the $M$-dimensional
 partial projection of this eigenprojection $P_b$.
 We define for
 a $k$-dimensional multi-index $M=(M_1,\cdots,M_k)$, $M_j \ge 1$ and
 $k=1,\cdots,N-1$,
$$
{\hat P} ^M_{k}= \left(I-\sum_{\sharp(b) = k}P_b^{M_k}\right)
\cdots \left(I-\sum_{\sharp(d) = 2} P_d^{M_{2}}\right)
(I-P^{M_1}),
$$
where note that $P^{M_1}=P_a^{M_1}=P_H^{M_1}$ for $\sharp(a)=1$ is
 uniquely determined. We also define for a $\sharp(b)$-dimensional
multi-index $M_{b} = (M_1, \cdots ,$ $ M_{\sharp(b) -1},
M_{\sharp(b)}) = ({\hat M}_{b}, M_{\sharp(b)})$
$$
{\tilde P}_{b}^{M_{b}}=P_b^{M_{\sharp(b)}}{\hat P}_{\sharp(b)
-1}^{{\hat M}_{b}}, \q 2\le\sharp(b)\le N.
$$
It is clear that
$$
\sum_{2\le\sharp(b)\le N} {\tilde P}^{M_{b}}_{b} = I - P^{M_1},
$$
provided that the component $M_k$ of $M_{b}$ depends only on the
 number $k$ but  not on  $b$.  In the following we use such 
$M_{b}$'s only. Under these circumstances, the following is known to
 hold.
\BP

\proclaim\nofrills
{\bf Theorem 1}\ {\rm{([En])}}. Let $N=n+1\ge 2$ and let
 $H_{N-1}=H_{n\ell}$ be the Hamiltonian for a local system
 $(H_{n\ell},\HH_{n\ell})$. Let suitable conditions on the
 decay rate for the pair potentials $V_{ij}(x_{ij})$ be 
satisfied {\rm (}see, e.g., Assumption 1 in 
{\rm{[Ki($N$)]}}{\rm )}. Let 
$\Vert |x^a|^2 P^M_a\Vert<\infty$ be satisfied for any integer 
$M\ge1$ and cluster decomposition $a$ with $2\le\sharp(a)\le N-1$.
 Let $f \in \HH^{N-1}$.  Then there  is a sequence 
$t_m \to \pm\infty \ (\text {as } m \to \pm\infty)$ and a sequence 
 $M^m_{b}$ of multi-indices whose components all tend to $\infty$ as
 $m \to \pm \infty$ such that for all cluster decompositions $b$, 
$2\le\sharp(b)\le N$, and 
$ \varphi \in C_0^\infty ({\Bbb R}^{3(\sharp(b) -1)}_{x_b})$
$$
\Vert \{ \varphi (x_b/t_m) - \varphi (v_b) \} 
{\tilde P}_{b}^{M^m_{b}}e^{-it_m H_{N-1}} f \Vert \to 0
\tag A
$$
as $m \to \pm\infty$.
\endproclaim

\BP

The asymptotic relation (A) roughly means that, if we restrict our
 attention to the part ${\tilde P}_{b}^{M^m_{b}}$ of the evolution 
$e^{-it H_{N-1}}f$, in which the particles inside any cluster of $b$
 are bounded while any two different clusters of $b$ are scattered,
 then the quantum-mechanical velocity $v_b=m_b^{-1}p_b$, where $m_b$
 is some diagonal mass matrix, is approximated by a classical value
 $v_b^{(c)}=\lim_{m\to\pm\infty}
(v_b{\tilde P}_{b}^{M^m_{b}}e^{-it_m H_{N-1}}f,
{\tilde P}_{b}^{M^m_{b}}e^{-it_m H_{N-1}}f)$
 asymptotically as $m\to\pm\infty$ and the local time $t$
 of the $N$ body system $H_{N-1}=H_{n\ell}$
 is asymptotically equal to the quotient of the configuration by the
 velocity of the scattered particles (or clusters, exactly 
speaking):
$$
\frac{\left|x_b\right|}{\left|v_b^{(c)}\right|}.\tag Q
$$
This means by $v_b=m_b^{-1}p_b$ that the local time $t$ is
 asymptotically and approximately
 measured if the values of the configurations and momenta for the
 scattered particles of the local system 
$(H_{N-1}, \HH_{N-1})=(H_{n\ell},\HH_{n\ell})$ are given.

\hyphenation{apparent}

We note that the time measured by (Q) is independent of the choice
 of cluster decomposition $b$ according to Theorem 1. This means that
 $t$ can be taken as a
 common parameter of motion inside the local system, and can be
 called {\it time} of the local system in accordance with the notion of
 `common time' in Newton's sense: ``relative, apparent, and common time,
 is some sensible and external (whether accurate or unequable)
 measure of duration by the means of motion, $\cdots$" ([New], p.6).
 Once we take 
$t$ as our notion of time for the system $(H_{n\ell},\HH_{n\ell})$,
 $t$ recovers the usual meaning of time, by the identity for
 $e^{-itH_{n\ell}}f$ known
 as the Schr\"odinger equation:
$$
\left(\frac{1}{i}\frac{d}{dt}+H_{n\ell}\right)e^{-itH_{n\ell}}f=0.
$$

\hyphenation{spaces}

Time $t=t_{(H_{n\ell},\HH_{n\ell})}$ is a notion defined only in
 relation with  the local system $(H_{n\ell},\HH_{n\ell})$. To other
 local system $(H_{mk},\HH_{mk})$, there is associated other local
 time $t_{(H_{mk},\HH_{mk})}$, and between 
$t=t_{(H_{n\ell},\HH_{n\ell})}$ and $t_{(H_{mk},\HH_{mk})}$,
 there is no relation, and they are completely independent notions.
 In other words, $\HH_{n\ell}$ and $\HH_{mk}$ are different spaces
 unless $n=m$ and $\ell=k$. And even when the two local systems 
$(H_{n\ell},\HH_{n\ell})$ and $(H_{mk},\HH_{mk})$ have a 
non-vanishing common part: $F_{n+1}^\ell\cap F_{m+1}^k\ne\emptyset$,
 the common part constitutes its own local system 
$(H_{pj},\HH_{pj})$, and its local time cannot be compared with
 those of the two bigger systems $(H_{n\ell},\HH_{n\ell})$ and 
$(H_{mk},\HH_{mk})$, because these three systems have different base
 spaces, Hamiltonians, and clocks. More concretely speaking, the
 times are measured through the quotients (Q) for each system. But
 the $L^2$-representations of the base Hilbert spaces 
$\HH_{n\ell},\HH_{mk},\HH_{pj}$ for those systems are different
 unless they are identical with each other, and the quotient (Q) has
 incommensurable meaning among these representations.

In this sense, local systems are independent mutually. Also they
 cannot be decomposed into pieces in the sense that the decomposed
 pieces constitute different local systems.

\BP

\subhead\nofrills{I.2. Relativity}
\endsubhead

\BP

\F
We note that the center of mass of a local system 
$(H_{n\ell},\HH_{n\ell})$ is always at the origin of the space
 coordinate system $x_{(H_{n\ell},\HH_{n\ell})}\in\R^3$ for the
 local system by the requirement: $\sum_{j\in F_{n+1}} m_j X_j=0$ in
 Axiom 2, and that the space coordinate system describes
 just the relative motions inside a local system
 by our formulation. The center of mass of a local system,
 therefore, cannot be identified from the local system
 itself, except that it is at the origin of the coordinates.

Moreover, just as we have seen in the previous section, we see
 that, not only the time coordinates $t_{(H_{n\ell},\HH_{n\ell})}$
 and $t_{(H_{mk},\HH_{mk})}$, but also the space coordinates 
$x_{(H_{n\ell},\HH_{n\ell})}\in \R^3$ and 
$x_{(H_{mk},\HH_{mk})}\in\R^3$ of these two local systems are
 independent mutually. Thus the space-time coordinates 
$(t_{(H_{n\ell},\HH_{n\ell})},x_{(H_{n\ell},\HH_{n\ell})})$ and 
$(t_{(H_{mk},\HH_{mk})},x_{(H_{mk},\HH_{mk})})$ are independent
 between two different local systems $(H_{n\ell},\HH_{n\ell})$ and 
$(H_{mk},\HH_{mk})$. In particular, insofar as the systems are
 considered as quantum-mechanical ones, there is no relation between 
their centers of mass. In other words, the center of mass of any 
local system cannot be identified by other local systems
 quantum-mechanically.

Summing these two considerations, we conclude: 
\roster
\item The center of mass of a local system $(H_{n\ell},\HH_{n\ell})$
 cannot be identified {\it quantum-mechanically} by any local system 
$(H_{mk},\HH_{mk})$ including the case 
$(H_{mk},\HH_{mk})=(H_{n\ell},\HH_{n\ell})$.

\item There is no {\it quantum-mechanical} relation between any two
 local coordinates 
$(t_{(H_{n\ell},\HH_{n\ell})},x_{(H_{n\ell},\HH_{n\ell})})$ and 
$(t_{(H_{mk},\HH_{mk})},x_{(H_{mk},\HH_{mk})})$ of two different
 local systems $(H_{n\ell},\HH_{n\ell})$ and $(H_{mk},\HH_{mk})$.
\endroster
\MP

\F
Utilizing these properties of the centers of mass and the
 coordinates of local systems, we may make any postulates concerning
\roster
\item the motions of the {\it centers of mass} of various local
 systems,
\endroster
 and
\roster
\item"(2)" the relation between two local coordinates of any
 two local systems.
\endroster
\MP

\F
In particular, we may impose {\it classical}
 postulates on them as far as the
 postulates are consistent in themselves. 
\MP

Thus we assume an arbitrary but fixed transformation:
$$
y_2=f_{21}(y_1)\tag Tr
$$
between the coordinate systems $y_j=(y_j^\mu)=(ct_j,x_j)$ for $j=1,2$,
 where $c$ is the speed of light in vacuum and $(t_j,x_j)$ is
 the space-time coordinates of the local system
 $L_j=(H_{n_j\ell_j},\HH_{n_j\ell_j})$. 
We regard these coordinates $y_j=(ct_j,x_j)$ as {\it classical}
coordinates, when we consider 
the motions of centers of mass and the relations of
 coordinates of various local systems.
We now postulate the general principle of relativity on
 the physics of the centers of mass:

\MP

\hyphenation{centers}

{\bf Axiom 4.} 
The laws of physics which control the relative motions of the
 centers of mass of local systems are covariant under the change of
 the coordinates from \linebreak
 $(ct_{(H_{mk},\HH_{mk})},x_{(H_{mk},\HH_{mk})})$ to 
$(ct_{(H_{n\ell},\HH_{n\ell})},x_{(H_{n\ell},\HH_{n\ell})})$
 of the reference frame local systems for any
 pair $(H_{mk},\HH_{mk})$ and $(H_{n\ell},\HH_{n\ell})$ of
 local systems.

\MP

We note that this axiom is
 consistent with the Euclidean metric adopted for the 
quantum-mechanical coordinates inside a local system, 
because Axiom 4 is concerned with classical motions of the centers
 of mass {\it outside} local systems, and we are dealing here with
 a different aspect of nature from the quantum-mechanical one 
{\it inside} a local system.

Axiom 4 implies the invariance of the distance under the change of
 coordinates between two local systems. Thus the metric tensor
 $g_{\mu\nu}(ct,x)$ which appears here satisfies the transformation rule:
$$
g^1_{\mu\nu}(y_1)=g^2_{\alpha\beta}(f_{21}(y_1))
\frac{\partial f^\alpha_{21}}{\partial y_1^\mu}(y_1)
\frac{\partial f_{21}^\beta}{\partial y_1^\nu}(y_1),\tag Me
$$
where $y_1=(ct_1,x_1)$; $y_2=f_{21}(y_1)$ is the transformation (Tr)
 in the above from $y_1=(ct_1,x_1)$ to $y_2=(ct_2,x_2)$; and 
$g_{\mu\nu}^j(y_j)$ is the metric tensor expressed in the classical
 coordinates $y_j=(ct_j,x_j)$ for $j=1,2$.

The second postulate is the principle of equivalence, which asserts
 that the classical coordinate system 
$(ct_{(H_{n\ell},\HH_{n\ell})},x_{(H_{n\ell},\HH_{n\ell})})$ is a
 local Lorentz system of coordinates, insofar as it is concerned
 with the classical behavior of the center of mass of the local
 system $(H_{n\ell},\HH_{n\ell})$:

\MP

{\bf Axiom 5.} 
The metric or the gravitational tensor $g_{\mu\nu}$ for the center
 of mass of a local system $(H_{n\ell},\HH_{n\ell})$ in the
 coordinates 
$(ct_{(H_{n\ell},\HH_{n\ell})},x_{(H_{n\ell},\HH_{n\ell})})$ of
 itself are equal to $\eta_{\mu\nu}$, where $\eta_{\mu\nu}=0$ for 
$\mu\ne\nu$, $=1$ for $\mu=\nu=1,2,3$, and $=-1$ for $\mu=\nu=0$.

\MP

Since, at the center of mass, the classical space
 coordinates $x=0$, Axiom 5 together with the transformation rule
 (Me) in the above yields
$$
g^1_{\mu\nu}(f^{-1}_{21}(ct_2,0))=\eta_{\alpha\beta}
\frac{\partial f^\alpha_{21}}{\partial y_1^\mu}(f^{-1}_{21}(ct_2,0))
\frac{\partial f_{21}^\beta}{\partial y_1^\nu}(f^{-1}_{21}(ct_2,0)).
\tag Equi
$$
Also by the same reason: $x=0$ at the center of mass, the relativistic
 proper time 
$d\tau=\sqrt{-g_{\mu\nu}(ct,0)dy^\mu dy^\nu}
=\sqrt{-\eta_{\mu\nu}dy^\mu dy^\nu}$ at the origin of a local system
 is equal to $c$ times the quantum-mechanical proper time $dt$
 of the system.

\BP

By the fact that the classical Axioms 4 and 5 of physics
 are imposed on the centers of mass which are uncontrollable 
 quantum-mechanically, and on the relation between the coordinates of
 different, therefore quantum-mechanically non-related local systems,
 the consistency of classical relativistic Axioms 4 and 5
 with quantum-mechanical Axioms 1--3 is clear:

\BP

\proclaim\nofrills
{\bf Theorem 2.}\ Axioms 1 to 5 are consistent.
\endproclaim


\BP

\subhead\nofrills{I.3. Observation}
\endsubhead

\BP

\F
Thus far, we did not mention any about the physics which
 is actually observed. We have just given two aspects of nature which
 are mutually independent. We will introduce a procedure to yield what
 we observe when we see nature. This procedure will not be
 contradictory with the two aspects of nature which we have discussed,
 as the procedure is concerned solely with ``{\it how nature looks, at
 the observer}," i.e. it is solely concerned with
 ``{\it at the place of the observer, how nature looks},"
 with some abuse of the word ``place."
 The validity of the procedure should be judged merely
 through the comparison between the observation and the prediction
 given by our procedure.

 We note that we can observe only a finite number of disjoint
 systems, say $L_1,\cdots,L_k$ with $k\ge1$ a finite integer. We
 cannot grasp an infinite number of systems at a time. Further each
 system $L_j$ must have only a finite number of elements by the same
 reason. Thus these systems $L_1,\cdots,L_k$ may be identified with
 local systems in the sense of section I.1.

Local systems are quantum-mechanical systems, and their coordinates
 are confined to their insides insofar as we appeal to Axioms 1--3.
 However we postulated Axioms 4 and 5 on the classical aspects of
 those coordinates, which make the local coordinates of a local
 system a classical reference frame for the centers of mass of other
 local systems. This leaves us the room to define observation as the
 {\it classical} observation of the centers of mass of local systems
 $L_1,\cdots,L_k$. We call this an observation of $L=(L_1,\cdots,L_k)$
 inquiring into sub-systems $L_1,\cdots,L_k$, where $L$ is a local
 system consisting of the particles which belong to one of the local
 systems $L_1,\cdots,L_k$.

When we observe the sub-local systems $L_1,\cdots,L_k$ of $L$,
 we observe the relations or motions among these sub-systems.
 Internally the local system $L$ behaves following the Hamiltonian
 $H_L$ associated to the local system $L$. However the actual
 observation differs from what the pure quantum-mechanical
 calculation gives for the system $L$. For example,
 when an electron is scattered by a nucleus with relative
 velocity close to that of light, the observation is different
 from the pure quantum-mechanical prediction.

In the usual explanation of this phenomenon, one introduces Dirac
 equation, and calculates differential cross section. However,
 the calculation only applies to that experiment or to the case
 which can be described by the Dirac equation, and no gravity is
 included.
\BP

We propose below a procedure which explains gravity as well
 as quantum-mechanical forces in one framework. 
\MP

The quantum-mechanical process inside the local system $L$ is
 described by the evolution
$$
\exp(-it_LH_L)f,
$$
when the initial state $f$ of the system and the local time $t_L$
 of the system are given. The Hamiltonian $H_L$ is decomposed as
 follows in virtue of the local Hamiltonians $H_1,\cdots,H_k$,
 which correspond to the sub-local systems $L_1,\cdots,L_k$:
$$
H_L=H^b+T+I,\quad H^b=H_1+\cdots+H_k.
$$
Here $b=(C_1,\cdots,C_k)$ is the cluster decomposition corresponding
 to the decomposition $L=(L_1,\cdots,L_k)$ of $L$; $H^b=H_1+\cdots+H_k$
 is the sum of the internal energies $H_j$ inside $L_j$, and is defined
 in the internal state space $\HH^b=\HH^b_1\otimes\cdots\otimes\HH^b_k$;
 $T=T_b$ denotes the intercluster free energy among the clusters
 $C_1,\cdots,C_k$ defined in the external state space $\HH_b$;
 and $I=I_b=I_b(x)=I_b(x_b,x^b)$ is the sum of the intercluster
 interactions between various two different clusters in the cluster
 decomposition $b$ (cf. the explanation after Definition 3 in section
 I.1).

The main concern in this process would be the case that the 
clusters $C_1,\cdots,C_k$ form asymptotically bound states as
 $t_L\to\infty$, since other cases are hard to be observed along
 the process if the observer's concern is upon the final state of
 the sub-systems $L_1,\cdots,L_k$.

The evolution $\exp(-it_LH_L)f$ behaves asymptotically as
 $t_L\to\infty$ as follows for some bound states $g_1,\cdots,g_k$
 ($g_j\in\HH^b_j$) of local Hamiltonians $H_1,\cdots,H_k$ and for
 some $g_0$ belonging to the external state space $\HH_b$:
$$
\exp(-it_LH_L)f\sim \exp(-it_L h_b)g_0
\otimes\exp(-it_LH_1)g_1\otimes\cdots\otimes \exp(-it_LH_k)g_k,
\quad k\ge 1,
\tag P
$$
where $h_b=T_b+I_b(x_b,0)$. It is easy to see that
 $g=g_0\otimes g_1\otimes\cdots\otimes g_k$ is given by
$$
g=g_0\otimes g_1\otimes\cdots\otimes g_k=\Omega_b^{+\ast} f
=P_b \Omega_b^{+\ast} f,
$$
provided that the decomposition of the evolution $\exp(-it_LH_L)f$
 is of the simple form as in (P). Here $\Omega_b^{+\ast}$ is the
 adjoint of a canonical wave operator ([De]) corresponding to the
 cluster decomposition $b$:
$$
\Omega_b^+=s{\text{-}}\lim_{t\to\infty}\exp(itH_L)\cdot
\exp(-ith_b)\otimes\exp(-itH_1)\otimes\cdots\otimes\exp(-itH_k)P_b,
$$
where $P_b$ is the eigenprojection onto the eigenspace of the
 Hamiltonian $H^b=H_1+\cdots+H_k$. The process (P) just describes
 the quantum-mechanical process inside the local system $L$, and
 does not specify any meaning related with observation up to the
 present stage.

To see what we observe at actual observations, let us reflect
a process of observation of scattering phenomena. We note that the
 observation of scattering phenomena is concerned with
 their initial and final stages by what the scattering itself means.
 At the final stage of observation of scattering processes,
 the quantities observed are firstly the points hit by the scattered
 particles on the screen stood against them. If the circumstances
 are properly set up, one can further indicate the momentum
 of the scattered particles at the final stage to the extent
 that the uncertainty principle allows. Consider, e.g.,
 a scattering process of an electron by a nucleus. Given the
 magnitude of initial momentum of an electron relative to the
 nucleus, one can infer the magnitude of momentum of the electron
 at the final stage as being equal to the initial one by the law
 of conservation of energy, since the electron and the nucleus are
 far away at the initial and final stages so that the potential
 energy between them can be neglected compared to the
 relative kinetic energy. The direction of momentum at the final
 stage can also be indicated, up to the error due to the
 uncertainty principle, by setting a sequence of slits toward
 the desired direction at each point on the screen so that the
 observer can detect only the electrons scattered to that direction.
 The magnitude of momentum at initial stage can be selected in
 advance by applying a uniform magnetic field to the electrons,
 perpendicularly to their momenta, so that they circulate around
 circles with the radius proportional to the magnitude of momentum,
 and then by setting a sequence of slits midst the stream of those
 electrons. The selection of magnitude of initial momentum
 makes the direction of momentum ambiguous due to the
 uncertainty principle, since the sequence of slits lets the
 position of electrons accurate to some extent.
 To sum up, the sequences of slits at the initial and final stages
 necessarily require to take into account the uncertainty
 principle so that some ambiguity remains in the observation.
 
However, in the actual observation of a {\it single} particle,
 we {\it have to decide} at which point on the screen the 
particle hits and which momentum the particle has, using
 the prepared apparatus like the sequence of slits located at each 
point on the screen. Even if we impose an interval for the 
observed values, we {\it have to assume} that the boundaries of 
the interval are sharply designated. These are the assumption 
which we always impose on ``observations" implicitly. I.e., 
we idealize the situation in any observation or in any measurement
 of a single particle so that the observed values for each particle
 are sharp for both of the configuration and momentum. In this 
sense, the values observed actually for each particle must be 
classical. We have then necessary and sufficient conditions
 to make predictions about the differential cross section, as we 
will see in subsection I.3.1.

 Summarizing, we observe just the classical quantities for each 
particle at the final stage of all observations. In other words, 
we have to  {\it presuppose} that the values observed for each 
particle have sharp values, even if we cannot know the values 
actually. We can apply to this fact the remark stated in the 
third paragraph of this section about the possibility of defining 
observation as that of the {\it classical} centers of mass of 
local systems, and may assume that the actually observed values 
follow the classical Axioms 4 and 5. Those sharp values actually 
observed for each particle give, when summed over the large 
number of particles, the probabilistic nature of physical 
phenomena, i.e. that of scattering phenomena.

Theoretically, the quantum-mechanical, probabilistic
 nature of scattering processes is described by differential
 cross section, defined as the square of the absolute value
 of the scattering amplitude obtained from scattering operators
 $S_{bd}=W_b^{+\ast} W_d^- $, where $W_b^\pm$ are usual wave
 operators. Given the magnitude of the initial momentum of the
 incoming particle and the scattering angle, the differential
 cross section gives a prediction about the probability at which
 point and to which direction on the screen each particle hits
 on the average. However, as we have remarked, the idealized 
point on the screen hit by each particle and the scattering angle 
given as an idealized difference between the directions of the 
initial and final momenta of each particle have sharp values, and 
the observation at the final stage is {\it classical}. We are 
then required to supplement these classical observations with 
taking into account the classical relativistic effects on those 
classical quantities, e.g., on the configuration and the momentum 
of each particle.
\BP

\F
{\bf I.3.1.} As the first step of the relativistic modification of the
 scattering process, we consider the scattering amplitude 
$\SS(E,\theta)$, where $E$ denotes the energy level of the scattering
 process and $\theta$ is a parameter describing the direction of the
 scattered particles. Following our remark made in the previous
 paragraph, we make the following postulate on the scattering
 amplitude observed in actual experiment:
\BP

{\bf Axiom 6.1.} When one observes the final stage of scattering
 phenomena, the total energy $E$ of the scattering process should be
 regarded as a classical quantity and is replaced by a relativistic
 quantity, which obeys the relativistic change of coordinates from
 the scattering system to the observer's system. 
\BP

Since it is not known much about $\SS(E,\theta)$ in the many body
 case, we consider an example of the two body case: Consider a
 scattering phenomenon of an electron by a Coulomb potential 
$Ze^2/r$, where $Z$ is a real number, $r=|x|$, and $x$ is the
 position vector of the electron relative to the scatterer. We assume
 that the mass of the scatterer is large enough compared to that of
 the electron and that $|Z|/137\ll 1$. Then quantum mechanics gives
 the differential cross section in a Born approximation:
$$
\frac{d\sigma}{d\Omega}=|{\Cal S}(E,\theta)|^2 
\approx \frac{Z^2e^4}{16E^2\sin^4(\theta/2)},
$$
where $\theta$ is the scattering angle and $E$ is the total
 energy of the system of the electron and the scatterer.
 We assume that the observer is stationary with respect to
 the center of mass of this system of an electron and the
 scatterer. Then, since the electron is far away from the
 scatterer after the scattering and the mass of the scatterer
 is much larger than that of the electron, we may suppose that
 the energy $E$ in the formula in the above can be replaced by the
 {\it classical} kinetic energy of the electron by Axiom 6.1.
 Then, assuming that the speed $v$ of the electron relative to
 the observer is small compared to the speed $c$ of light 
in vacuum and denoting the rest mass of the electron by $m$,
 we have by Axiom 6.1 that $E$ is observed to have the 
following relativistic value:
$$
E'=c\sqrt{p^2+m^2c^2}-mc^2
=\frac{mc^2}{\sqrt{1-(v/c)^2}}-mc^2\approx 
\frac{mv^2}{2\sqrt{1-(v/c)^2}},
$$
where $p=mv/\sqrt{1-(v/c)^2}$ is the relativistic momentum of
 the electron. Thus the differential cross section should be
 observed approximately equal to
$$
\frac{d\sigma}{d\Omega}
\approx \frac{Z^2e^4}{4m^2v^4\sin^4(\theta/2)}(1-(v/c)^2).\tag RDC
$$
This coincides with the usual relativistic prediction obtained
 from the Klein-Gordon equation by a Born approximation. See 
[Ki, p.297] for a case which involves the spin of the electron.

Before proceeding to the inclusion of gravity in the general $k$ cluster
 case, we review this two body case. We note that the two body case
 corresponds to the case $k=2$, where $L_1$ and $L_2$ consist of
 single particle, therefore the corresponding Hamiltonians $H_1$
 and $H_2$ are zero operators on $\HH^0={\Bbb C}=$ the complex
 numbers. The scattering amplitude ${\Cal S}(E,\theta)$ in this 
case is an integral kernel of the scattering matrix 
${\hat S}={\Cal F}S{\Cal F}^{-1}$, where $S=W^{+\ast}W^-$ is
 a scattering operator; 
$W^\pm=s$-$\lim_{t\to\pm\infty}\exp(itH_L)\exp(-itT)$ are
 wave operators ($T$ is negative Laplacian for short-range
 potentials under an appropriate unit system, while it has
 to be modified when long-range potentials are included);
 and $\Cal F$ is Fourier transformation so that 
${\Cal F}T{\Cal F}^{-1}$ is a multiplication operator by 
$|\xi|^2$ in the momentum representation $L^2({\Bbb R}^3_\xi)$.
 By definition, $S$ commutes with $T$. This makes ${\hat S}$
 decomposable with respect to $|\xi|^2={\Cal F}T{\Cal F}^{-1}$:
 For {\it a.e.} $E>0$, there is a unitary operator ${\Cal S}(E)$
 on $L^2(S^2)$, $S^2$ being two dimensional sphere with radius
 one, such that for {\it a.e.} $E>0$ and $\omega\in S^2$
$$
({\hat S}h)(\sqrt{E}\omega)
=\left({\Cal S}(E)h(\sqrt{E}\cdot)\right)(\omega),
\quad h\in L^2({\Bbb R}^3_\xi)
=L^2((0,\infty),L^2(S^2_\omega),|\xi|^2d|\xi|).
$$
Thus ${\hat S}$ can be written as ${\hat S}=\{{\Cal S}(E)\}_{E>0}$.
 It is known [I-Ki] that ${\Cal S}(E)$ can be expressed as 
$$
({\Cal S}(E)\varphi)(\theta)=\varphi(\theta)
-2\pi i \sqrt{E}\int_{S^2}
{\Cal S}(E,\theta,\omega)\varphi(\omega)d\omega
$$
for $\varphi\in L^2(S^2)$. The integral kernel 
${\Cal S}(E,\theta,\omega)$ with $\omega$ being 
the direction of initial wave, is the scattering amplitude
 ${\Cal S}(E,\theta)$ stated in the above and
 $|{\Cal S}(E,\theta,\omega)|^2$ is called differential
 cross section. These are the most important quantities in physics
 in the sense that they are the {\it only} quantities which can be
 observed in actual physical observation. 

The energy level $E$ in the previous example thus corresponds
 to the energy shell $T=E$, and the replacement of $E$ by $E'$
 in the above means that $T$ is replaced by a {\it classical
 relativistic} quantity $E'=c\sqrt{p^2+m^2c^2}-mc^2$. We have then
 seen that the calculation in the above gives a correct relativistic
 result, which explains the actual observation.

Axiom 6.1 is concerned with the observation of the final stage
 of scattering phenomena. To include the gravity into our
 consideration, we extend Axiom 6.1 to the intermediate process
 of quantum-mechanical evolution. The intermediate process cannot
 be an object of any {\it actual} observation, because the
 intermediate observation would change the process itself,
 consequently the result observed at the final stage would be
 altered. Our next Axiom 6.2 is an extension of Axiom 6.1 from
 the {\it actual} observation to the {\it ideal} observation in
 the sense that Axiom 6.2 is concerned with such invisible
 intermediate processes and modifies the {\it ideal} intermediate
 classical quantities by relativistic change of coordinates. The
 spirit of the treatment developed below is to trace the
 quantum-mechanical paths by ideal observations so that the
 quantities will be transformed into classical quantities at
 each step, but the quantum-mechanical paths will not be altered
 owing to the {\it ideality} of the observations. The classical
 Hamiltonian obtained at the last step will be ``requantized" to
 recapture the quantum-mechanical nature of the process, therefore
 the ideality of the intermediate observations will be realized
 in the final expression of the propagator of the observed system.

\BP

\F
{\bf I.3.2.} With these remarks in mind, we return to the general
 $k$ cluster case, and consider a way to include gravity
 in our framework.

In the scattering process into $k\ge1$ clusters, what we observe
 are the centers of mass of those $k$ clusters $C_1,\cdots,C_k$,
 and of the combined system $L=(L_1,\cdots,L_k)$. In the example
 of the two body case of the previous subsection, only the combined
 system $L=(L_1,L_2)$ appears due to $H_1=H_2=0$, therefore the
 replacement of $T$ by $E'$ is concerned with the free energy
 between two clusters $C_1$ and $C_2$ of the combined system 
$L=(L_1,L_2)$. 

Following this treatment of $T$ in subsection I.3.1, we
 replace $T=T_b$ in the exponent of 
$\exp(-it_Lh_b)=\exp(-it_L(T_b+I_b(x_b,0)))$ on the right hand
 side of the asymptotic relation (P) by the relativistic kinetic
 energy $T'_b$ among the clusters $C_1,\cdots,C_k$ around the
 center of mass of $L=(L_1,\cdots,L_k)$, defined by
$$
T'_b=\sum_{j=1}^k\left(c\sqrt{p_j^2+m_j^2c^2}-m_jc^2\right).
\tag Energy 1
$$
Here $m_j>0$ is the rest mass of the cluster $C_j$, which involves
 all the internal energies like the kinetic energies inside $C_j$
 and the rest masses of the particles inside $C_j$, and $p_j$ is
 the relativistic momentum of the center of mass of $C_j$ inside
 $L$ around the center of mass of $L$. For simplicity, we assume
 that the center of mass of $L$ is stationary relative to the
 observer. Then we can set in the exponent of 
$\exp(-it_L(T'_b+I_b(x_b,0)))$
$$
t_L=t_O,\tag Time 1
$$
where $t_O$ is the observer's time.

 For the factors $\exp(-it_LH_j)$ on the right hand side of (P),
 the object of the ({\it ideal}) observation is the centers of
 mass of the $k$ number of clusters $C_1,\cdots,C_k$. These are
 the ones which now require the relativistic treatment. Since we
 identify the clusters $C_1,\cdots,C_k$ as their centers of mass
 moving in a classical fashion, $t_L$ in the exponent of
 $\exp(-it_LH_j)$ should be replaced by $c^{-1}$ times
 the classical relativistic proper time at the origin of the local
 system $L_j$, which is equal to the quantum-mechanical local time
 $t_j$ of the sub-local system $L_j$. By the same reason and by the
 fact that $H_j$ is the internal energy of the cluster $C_j$
 relative to its center of mass, it would be justified to
 replace the Hamiltonian $H_j$ in the exponent of $\exp(-it_jH_j)$
 by the classical relativistic energy {\it inside} the cluster
 $C_j$ around its center of mass
$$
H'_j=m_jc^2, \tag Energy 2
$$
where $m_j>0$ is the same as in the above.

Summing up, we arrive at the following postulate, which has the
 same spirit as in Axiom 6.1 and includes Axiom 6.1 as a special
 case concerned with actual observation:
\BP

{\bf Axiom 6.2.} In either actual or ideal observation, the
 space-time coordinates \linebreak $(ct_L,x_L)$ and the four momentum
 $p=(p^\mu)=(E_L/c,p_L)$ of the observed system $L$ should be
 replaced by classical relativistic quantities, which are transformed
 into the classical quantities $(ct_O,x_O)$ and $p=(E_O/c,p_O)$ in
 the observer's system $L_O$ according to the relativistic change
 of coordinates specified in Axioms 4 and 5. Here $t_L$ is the local
 time of the system $L$ and $x_L$ is the internal space coordinates
 inside the system $L$; and $E_L$ is the internal energy of the system
 $L$ and $p_L$ is the momentum of the center of mass of the system $L$.
\BP

In the case of the present scattering process into $k$ clusters,
 the system $L$ in this axiom is each of the local systems $L_j$
 $(j=1,2,\cdots,k)$ and $L$.

We continue to consider the $k$ centers of mass of the clusters
 $C_1,\cdots,C_k$. At the final stage of the scattering process,
 the velocities of the centers of mass of the clusters
 $C_1,\cdots,C_k$ would be steady, say $v_1,\cdots,v_k$, 
relative to the observer's system. Thus, according to Axiom 6.2, 
the local times $t_j$ $(j=1,2,\cdots,k)$ in the exponent of 
$\exp(-it_jH'_j)$, which are equal to $c^{-1}$ times the
 relativistic proper times at the origins $x_j=0$ of the
 local systems $L_j$, are expressed in the observer's time
 coordinate $t_O$ by
$$
t_j=t_O\sqrt{1-(v_j/c)^2}\approx t_O\left(1-v_j^2/(2c^2)\right),
\quad j=1,2,\cdots,k,\tag Time 2
$$
where we have assumed $|v_j/c|\ll 1$ and used Axioms 4 and 5 
to deduce the Lorentz transformation:
$$
t_j=\frac{t_O-(v_j/c^2)x_O}{\sqrt{1-(v_j/c)^2}},\quad 
x_j=\frac{x_O-v_jt_O}{\sqrt{1-(v_j/c)^2}}.
$$
(For simplicity, we wrote the Lorentz transformation for the
 case of 2-dimensional space-time.)

Inserting (Energy 1-2) and (Time 1-2) into the right-hand side of
 (P), we obtain a classical approximation of the evolution:
$$
\exp\left(-it_O[(T'_b+I_b(x_b,0)+H'_1+\cdots+H'_k)
-(m_1v_1^2/2+\cdots+m_kv_k^2/2)]\right)\tag AP
$$
under the assumption that  $|v_j/c|\ll 1$ for all $j=1,2,\cdots,k$.

 What we want to clarify is the final stage of the scattering
 process. Thus as we have mentioned, we may assume that
 all clusters $C_1,\cdots,C_k$ are far
 away from any of the other clusters and moving almost in steady
 velocities $v_1,\cdots,v_k$ relative to the observer. We denote by
 $r_{ij}$ the distance between two centers of mass of the clusters
 $C_i$ and $C_j$ for $1\le i<j\le k$. Then, according to our spirit
 that we are observing the behavior of the centers of mass of the
 clusters $C_1,\cdots,C_k$ in {\it classical} fashion following Axioms
 4 and 5, the clusters $C_1,\cdots,C_k$ can be regarded to have
 gravitation among them. This gravitation can be calculated if we
 assume Einstein's field equation, $|v_j/c|\ll 1$, and certain
 conditions that the gravitation is weak (see [M, section 17.4]),
 in addition to our Axioms 4 and 5. As an approximation of the 
 first order, we obtain the gravitational potential of Newtonian
 type for, e.g., the pair of the clusters 
$C_1$ and $U_1=\bigcup_{i=2}^kC_i$:
$$
-G\sum_{i=2}^km_1m_i/r_{1i},
$$
where $G$ is Newton's gravitational constant.

Considering the $k$ body classical problem for the $k$ clusters
 $C_1,\cdots,C_k$ moving in the sum of these gravitational fields,
 we see that the sum of the kinetic energies of $C_1,\cdots,C_k$
 and the gravitational potentials among them is constant by the
 classical law of conservation of energy:
$$
m_1v_1^2/2+\cdots+m_kv_k^2/2-G\sum_{1\le i<j\le k}m_im_j/r_{ij}
={\text{constant}}.
$$
Assuming that $v_j\to v_{j\infty}$ as time tends to infinity, we
 have constant $=m_1v_{1\infty}^2/2+\cdots+m_kv_{k\infty}^2/2$.
 Inserting this relation into (AP) in the above, we obtain the
 following as a classical approximation of the evolution (P):
$$
\exp\left(-it_O\left[T'_b+I_b(x_b,0)
+\sum_{j=1}^k(m_jc^2-m_jv_{j\infty}^2/2)
-G\sum_{1\le i<j\le k}m_im_j/r_{ij}\right]\right).
\tag CP
$$
What we do at this stage are {\it ideal} observations, and these
 observations should not give any sharp classical values. Thus
 we have to consider (CP) as a {\it quantum-mechanical evolution}
 and we have to recapture the quantum-mechanical feature of the
 process. To do so we replace $p_j$ in $T'_b$ in (CP) by a
 quantum-mechanical momentum $D_j$, where $D_j$ is a differential
 operator $-i\frac{\partial}{\partial x_j}
=-i\left(\frac{\partial}{\partial x_{j1}},
\frac{\partial}{\partial x_{j2}},
\frac{\partial}{\partial x_{j3}}\right)$ with respect 
to the 3-dimensional coordinates $x_j$ of the center of mass
 of the cluster $C_j$. Thus  the actual process should be
 described by (CP) with $T'_b$ replaced by a quantum-mechanical
 Hamiltonian
$$
\tT_b=\sum_{j=1}^k\left(c\sqrt{D_j^2+m_j^2c^2}-m_jc^2\right).
$$
This procedure may be called ``requantization," and is summarized
 as the following axiom concerning the ideal observation.
\BP

{\bf Axiom 6.3.} In the expression describing the classical process
 at the time of the {\it ideal} observation, the intercluster momentum
 $p_j=(p_{j1},p_{j2},p_{j3})$ should be replaced by a quantum-mechanical
 momentum $D_j=-i\left(\frac{\partial}{\partial x_{j1}},
\frac{\partial}{\partial x_{j2}},\frac{\partial}{\partial x_{j3}}\right)$.
 Then this gives the evolution describing the intermediate 
{\it quantum-mechanical} process.

\BP

We thus arrive at an approximation for a quantum-mechanical
 Hamiltonian including gravitational effect up to a constant term,
 which depends on the system $L$ and its decomposition into
 $L_1,\cdots,L_k$, but not affecting the quantum-mechanical
 evolution, therefore can be eliminated:
$$
\aligned
\tH_L&=\tT_b+I_b(x_b,0)-G\sum_{1\le i<j\le k}m_im_j/r_{ij}\\
&=\sum_{j=1}^k\left(c\sqrt{D_j^2+m_j^2c^2}-m_jc^2\right)+I_b(x_b,0)
-G\sum_{1\le i<j\le k}m_im_j/r_{ij}.
\endaligned
\tag QMG
$$
 We remark that the gravitational terms here come from the substitution
 of local times $t_j$ to the time $t_L$ in the factors $\exp(-it_L H_j)$
 on the right-hand side of (P). This form of Hamiltonian in (QMG) is
 actually used in [Li] with $I_b=0$ to explain the stability and
 instability of cold stars of large mass, showing the effectiveness
 of the Hamiltonian.

 Summarizing these arguments from (P) to (QMG), we have obtained
 the following {\it interpretation} of the observation of the
 quantum-mechanical evolution: To get our prediction for the
 observation of local systems $L_1,\cdots,L_k$, the
 quantum-mechanical evolution of the combined local system
 $L=(L_1,\cdots,L_k)$
$$
\exp(-it_LH_L)f
$$
should be replaced by the following evolution, in the
 approximation of the first order under the assumption that
 $|v_j/c|\ll 1$ $(j=1,2,\cdots,k)$ and the gravitation is weak,
$$
(\exp(-it_O\tH_L)\otimes 
\underbrace{I\otimes\cdots\otimes I}_{k \ \text{factors}})P_b
\Omega_b^{+\ast}f,
\tag GP
$$
provided that the original evolution $\exp(-it_LH_L)f$ decomposes
 into $k$ number of clusters $C_1,\cdots,C_k$ as $t_L\to\infty$ in
 the sense of (P). Here $b$ is the cluster decomposition
 $b=(C_1,\cdots,C_k)$ that corresponds to the decomposition
 $L=(L_1,\cdots,L_k)$ of $L$; $t_O$ is the observer's time; and
$$
\tH_L={\tilde T}_b+I_b(x_b,0)-G\sum_{1\le i<j\le k}m_im_j/r_{ij}
 \tag RH
$$
is the relativistic Hamiltonian inside $L$ given by (QMG), which
 describes the motion of the centers of mass of the clusters
 $C_1,\cdots,C_k$. 

We remark that (GP) may produce a bound state combining 
$C_1,\cdots,C_k$ as $t_O\to\infty$ therefore for all $t_O$, 
due to the gravitational potentials in the exponent. Note that
 this is not prohibited by our assumption that $\exp(-it_LH_L)f$
 has to decompose into $k$ clusters $C_1,\cdots,C_k$, because the
 assumption is concerned with the original Hamiltonian $H_L$ but
 not with the resultant Hamiltonian $\tH_L$.
\BP

\hyphenation{dif-fer-en-tial}
\hyphenation{small}
\hyphenation{since}
\hyphenation{stated}

 Extending our primitive assumption Axiom 6.1, which was valid for
 an example stated in subsection I.3.1, we have arrived at a
 relativistic Hamiltonian $\tH_L$, which would describe approximately
 the intermediate process, under the assumption
 that the gravitation is weak and the velocities of the particles
 are small compared to $c$, by using the Lorentz transformation.
 We note that, since we started our argument from the asymptotic
 relation (P), which is concerned with the final stage of
 scattering processes, we could assume that the velocities
 of particles are almost steady relative to the observer
 in the correspondent
 classical expressions of the processes, therefore we could appeal to
 the Lorentz transformations when performing the change of coordinates
 in the relevant arguments. 

 The final values of scattering amplitude should
 be calculated by using the Hamiltonian $\tH_L$. Then they would
 explain actual observations. This is our prediction for the observation
 of relativistic quantum-mechanical phenomena including the effects by
 gravity and quantum-mechanical forces.

 In the example discussed in subsection I.3.1, this approach gives
 the same result as (RDC) in the approximation of the first order,
 showing the consistency of our spirit. This can be seen by a
 representation formula of the scattering matrix similar to (3.7)
 in [I-Ki] for the present Hamiltonian $\tH_L={\tilde T}+V-GmM/r$
 with $V=Ze^2/r$ and $M$ being the mass of the scatterer: Define 
$\mu=(E+mc^2)^2/c^2-m^2c^2=c^{-2}E(E+2mc^2)$ for $E>0$, and set
$$
({\Cal F}_0(E)f)(\omega)=c^{-1}\sqrt{2(E+mc^2)}\mu^{1/4}
({\Cal F}f)(\sqrt{\mu}\omega),\quad f\in C_0^\infty({\Bbb R}^3),
$$
where ${\Cal F}$ is Fourier transformation, so that 
${\Cal F}_0(E)$ decomposes ${\tilde T}=c\sqrt{D^2+m^2c^2}-mc^2$
$$
{\Cal F}_0(E){\tilde T}f= E{\Cal F}_0(E)f,
$$
and satisfies
$$
\int_0^\infty\Vert {\Cal F}_0(E)f \Vert^2_{L^2(S^2)}dE
=\Vert f\Vert^2_{L^2({\Bbb R}^3)}.
$$
The scattering matrix ${\Cal S}(E)$ is then given as follows,
 in Born approximation of the first order under the assumption
 that $|Z|/137\ll 1$, as in (3.7) of [I-Ki]:
$$
{\Cal S}(E)\approx I-2\pi i{\Cal F}_0(E){\tilde V}{\Cal F}_0(E)^\ast,
$$
where ${\tilde V}$ is a modified Coulomb potential obtained from
 $V=Ze^2/r$ in accordance with the long-range tail of $V$, and
 we omitted the gravitational potential in $\tH_L$, since it is
 small compared to $V$. Then, calculating by using oscillatory
 integrals, we obtain the differential cross section
 $|{\Cal S}(E,\theta)|^2$ equal to (RDC), if we replace the
 quantum-mechanical quantity ${\tilde T}=E$ by the corresponding
 classical quantity 
$E'=c\sqrt{p^2+m^2c^2}-mc^2\approx mv^2/\{2\sqrt{1-(v/c)^2}\}$,
 assuming that the speed $v$ of the electron is small compared to $c$.

\hyphenation{physics}

We remark that our stand does not require the resultant Hamiltonian
 $\tH_L$ to satisfy the Lorentz invariance or other kinds of
 invariance under transformations among coordinate systems, unlike
 the usual attempts require in constructing relativistic quantum
 theories. We have just given a procedure to predict what we
 actually observe, but did not propose a physical law. Usual
 attempts identify physics with observation, and require such
 kind of invariance of observation. We separate observation from
 physics, allowing asymmetry to observation, but with preserving
 two mutually incompatible invariances for physics: Galilei
 invariance for internal quantum mechanics and general relativistic
 invariance for external classical physics. This becomes possible
 by our position that relativity is concerned with the external
 world outside local systems, but not with the internal physics,
 which is ruled by quantum mechanics. In fact, we postulated
 relativity as concerned with the centers of mass of local
 systems in Axioms 4 and 5, and in Axioms 6.1 and 6.2
 we clarified the role the relativity plays when observing
 the centers of mass. We refer the reader to [Ki-Fl] for further
 philosophical position of ours.

\BP

\vskip18pt

\centerline{\bf Part II. Examples}

\BP

\vskip12pt

\F
In this Part II we consider two examples of human size and of
 cosmological size following the spirit of the previous Part,
 both of which involve the quantum-mechanical aspects and
 relativistic aspects simultaneously.
\BP


\subhead\nofrills{II.1. Scattering of one neutron in
 a uniform gravitational field}
\endsubhead
\BP

\F
Consider the experiment done by Collela et al. [Co] of measuring
 the interference of one neutron. This experiment is described in
 some simplification as in the following \linebreak Figure 1:

\

\vskip6truemm

\font\lfont=line10

\define\righthd{$\vcenter{  \hbox{\lfont\char'00}}$}

\hskip3.88cm C \hskip3.88cm  D
\nopagebreak

\hskip4cm\righthd
\hskip-2mm---\hskip-0.4mm---\hskip-0.4mm---\hskip-0.4mm---
\hskip-2mm---\hskip-0.4mm---\hskip-0.6mm$\to$\hskip-0.6mm---
\hskip-2mm---\hskip-0.4mm---\hskip-0.4mm---\hskip-0.4mm---
\hskip-2mm---\hskip-2.12mm\righthd$\hskip-4mm\longrightarrow$
 O: Observer
\nopagebreak
\vskip-2.6mm
\nopagebreak
\hskip4.14cm $|$\hskip3.76cm $|$
\nopagebreak
\vskip-2.3mm
\nopagebreak
\hskip4.14cm $|$\hskip3.76cm $|$
\nopagebreak
\vskip-2.3mm
\nopagebreak
\hskip4.14cm $|$\hskip3.76cm $|$
\nopagebreak
\vskip-2.3mm
\nopagebreak
\hskip4.1cm $\uparrow$\hskip36.86mm $\uparrow$\hskip0.8cm
 the height BD $=L$
\nopagebreak
\vskip-2.3mm
\nopagebreak
\hskip4.14cm $|$\hskip3.76cm $|$
\nopagebreak
\vskip-2.3mm
\nopagebreak
\hskip4.14cm $|$\hskip3.76cm $|$
\nopagebreak
\vskip-2.3mm
\nopagebreak
\hskip4.14cm $|$\hskip3.76cm $|$
\nopagebreak
\vskip-2.62mm
\nopagebreak
\hskip3.1cm S $\longrightarrow$\hskip-1mm---\hskip-2mm\righthd
\hskip-2mm---\hskip-0.4mm---\hskip-0.4mm---\hskip-0.4mm---
\hskip-2mm---\hskip-0.4mm---\hskip-0.6mm$\to$\hskip-0.6mm---
\hskip-2mm---\hskip-0.4mm---\hskip-0.4mm---\hskip-0.4mm---\hskip-1.2mm
---\hskip-1.8mm\righthd
\nopagebreak

\hskip4.26cm A \hskip3.5cm B
\nopagebreak

\hskip6.02cm $\downarrow$
\nopagebreak

\hskip5.5cm EARTH
\nopagebreak

\vskip2mm
\nopagebreak

\hskip5.5cm Figure 1

\vskip6truemm

\F
A neutron beam emitted at S is split into two beams by an
 interferometer at A, and the two beams are recombined at point
 D by other interferometers or mirrors B and C. The height $L$
 of the line BD on the earth can be varied. The dependence on
 $L$ of the relative phase difference is given as follows,
according to the experiment of [Co], up to the error of about 1 \%:
$$
\hbar^{-1}mgLT,
$$
where $m$ is the mass of the neutron, $g$ is the acceleration by
 gravity, and $T$ is the (observed) time that the beams travel from
 C to D or A to B. This experiment shows that quantum mechanics and
 gravity play important roles {\it simultaneously} in the size of
 desktop environment. In fact, the lengths of the lines AB and BD are
 less than 10 cm in [Co].

This experiment can be explained in our context, if we see it
 as a 3-body scattering process of a neutron N by two mirrors
 B and C as in Figure 2.
\pagebreak

\

\vskip4truemm

\define\MARU#1{{\rm\ooalign{\hfil\lower.168ex\hbox{#1}
\hfil\crcr\mathhexbox20D}}}

\hskip3.88cm C \hskip3.88cm  D
\nopagebreak

\hskip4cm\righthd\hskip38mm
\hskip4mm O: Observer
\nopagebreak
\vskip-2.4mm
\nopagebreak
\hskip3.76cm $\ $
\vskip-2.3mm
\nopagebreak
\hskip3.76cm $\ $
\vskip-2.3mm
\nopagebreak
\hskip3.76cm $\ $
\vskip-2.3mm
\nopagebreak
\hskip3.76cm $\ $
\vskip-2.3mm
\nopagebreak
\hskip3.76cm $\ $
\vskip-2.3mm
\nopagebreak
\hskip5.14cm \MARU{\ } \hskip2mm N
\vskip-2.3mm
\nopagebreak
\hskip3.76cm $\ $
\vskip-2.3mm
\nopagebreak
\hskip3cm S \hskip6mm\hskip4truemm\hskip38mm\righthd
\vskip-3mm
\nopagebreak
\hskip4.24cm A \hskip3.8cm B
\nopagebreak

\hskip6.02cm $\downarrow$
\nopagebreak

\hskip5.5cm EARTH
\vskip2mm
\nopagebreak
\hskip5.5cm Figure 2

\vskip6truemm

\F
We denote the local system of the three bodies N, B, and C by $L$.
Let the masses of mirrors B and C be $M$, the neutron mass be $m$,
 and assume $0<m\ll M$. Let $x$, $X_B$ and $X_C$ denote the
 3-dimensional coordinates of N, B and C.
Then the Hamiltonian of this system $L$ with $\hbar=1$ is
$$
H=\frac{D_x^2}{2m}+\frac{D_B^2}{2M}+\frac{D_C^2}{2M},
$$
where $D_x$, $D_B$ and $D_C$ are the momentum operators
 $\frac{1}{i}\frac{\partial}{\partial x}
=\frac{1}{i}\left( \frac{\partial}{\partial x_1},
\frac{\partial}{\partial x_2},\frac{\partial}{\partial x_3}\right)$,
 $\frac{1}{i}\frac{\partial}{\partial X_B}$, and
 $\frac{1}{i}\frac{\partial}{\partial X_C}$
for N, B and C.
To separate the center of mass, we introduce the two sets of
 Jacobi coordinates:
$$
\left\{
\aligned
x^{(1)}&=x-X_C,\\
y^{(1)}&=X_B-\frac{mx+MX_C}{m+M},
\endaligned
\right.
\tag C1
$$
and
$$
\left\{
\aligned
x^{(2)}&=x-X_B,\\
y^{(2)}&=X_C-\frac{mx+MX_B}{m+M}.
\endaligned
\right.
\tag C2
$$
These choices of Jacobi coordinates $(x^{(j)}, y^{(j)})$ $(j=1,2)$
 correspond to two cluster decompositions $b^{(j)}=(C_1^{(j)},C_2^{(j)})$
 of $L$ such that $C_1^{(1)}=\{ \text{N, C}\}$ and 
$C_2^{(1)}=\{\text{B}\}$,
 or $C_1^{(2)}=\{ \text{N, B}\}$ and $C_2^{(2)}=\{ \text{C}\}$. In either
 case, $x^{(j)}$ is the internal coordinate inside the cluster
 $C_1^{(j)}$, and $y^{(j)}$ is the intercluster coordinate between
 two clusters $C_1^{(j)}$ and $C_2^{(j)}$.

Using these coordinates, we remove the center of mass of the system
 $L$. Then we obtain the Hamiltonian $H$ which has the same form for
 both coordinates:
$$
\split
&H=H^{(1)}+T^{(1)}=H^{(2)}+T^{(2)}, \\
&H^{(j)}=\frac{(D_x^{(j)})^2}{2\mu},
\quad   T^{(j)}=\frac{(D_y^{(j)})^2}{2\nu}.
\endsplit
$$
Here $D_x^{(j)}=\frac{1}{i}\frac{\partial}{\partial x^{(j)}}$ and
$D_y^{(j)}=\frac{1}{i}\frac{\partial}{\partial y^{(j)}}$ are
 momentum operators conjugate to $x^{(j)}$ and $y^{(j)}$
 $(j=1,2)$, and $\mu, \nu$ are the reduced masses:
$$
\mu^{-1}=m^{-1}+M^{-1}, \quad \nu^{-1}=M^{-1}+(m+M)^{-1}.
$$
Note that the operators $D_x^{(j)}$ and $D_y^{(j)}$ are mutually
 independent,
therefore $H^{(j)}$ commutes with $T^{(j)}$.
Thus the propagation
 of the 3-body system $L$ is given by
$$
\exp(-itH)f=\exp(-itT^{(j)})\exp(-itH^{(j)})f,\tag E
$$
where $f=f(x^{(j)}, y^{(j)})$ is the initial wave function at time
 $t=0$, just after the neutron has been split into two beams by the
 interferometer A. Here the {\it time} $t$ is the local time determined
 by the Hamiltonian $H$ or the correspondent local system $L$.

  $x^{(j)}$ is the distance vector between N and C, or between
 N and B, and $y^{(j)}$ is the distance vector between B and the
 center of mass of the system N+C, or between C and the center of
 mass of the system N+B. Therefore, as seen from the formula for
 $y^{(j)}$ in (C1) or (C2), we may regard it as
$$
y^{(1)}=X_B-X_C\quad \text{or}\quad y^{(2)}=X_C-X_B,
$$
for $M$ is larger enough than $m$.
We can thus regard
$y^{(j)}$ as constant during the scattering process, hence
$f(x^{(j)}, y^{(j)})$ can be regarded as a function of $x^{(j)}$ only.
(Exactly speaking, $f(x^{(j)},y^{(j)})$ can be written as
 $f(x^{(j)})F(y^{(j)})$ with $F(y^{(j)})$ close to the delta function
 $\delta_{\pm BC}$ having support at $y^{(j)}=X_B-X_C$ or
 $y^{(j)}=X_C-X_B$. But as we will see, the factor $F(y^{(j)})$
 does not play any essential role in our argument, and we can simply
 omit it from $f(x^{(j)},y^{(j)})$.)

Namely $f(x^{(j)}, y^{(j)})$ can be regarded as the wave function of
 the neutron N, and is split into two wave packets
$f_1(x^{(j)})$, $f_2(x^{(j)})$ at time $t=0$ by the interferometer A:
$$
f=f_1+f_2.
$$
$f_1$ is the wave packet moving to the direction from A to C, and
$f_2$ is the one from A to B. (E) can then be rewritten as follows:
$$
\exp(-itH)f=\exp(-itT^{(1)})\exp(-itH^{(1)})f_1+\exp(-itT^{(2)})
\exp(-itH^{(2)})f_2.
$$
As remarked in the above, we can regard
 $y^{(1)}=X_B-X_C$ or $y^{(2)}=X_C-X_B$,
 therefore we may set $T^{(1)}=T^{(2)}=T$. We thus have
$$
\exp(-itH)f= \exp(-itT) \{\exp(-itH^{(1)})f_1+\exp(-itH^{(2)})f_2\}.
\tag E1
$$
The description up to here is by the local time $t$ determined by
 the local system $L$.

$H^{(1)}$ is the Hamiltonian of the local system consisting of N and
 C, and the center of mass of N and C is regarded, by $m\ll M$, as
 located at C with the same height as the observer. Hence,
 corresponding to (Time 1) in subsection I.3.2, we can set in the first
 term $\exp(-itH^{(1)})f_1$ of (E1):
$$
t=t_O.
$$

$H^{(2)}$ is the Hamiltonian consisting of N and B, and its center of
 mass is regarded as located at B by $m\ll M$. Therefore that local
 system has a lower gravitational potential in amount $gL$ compared
 to the observer O, hence the local time $t$ of the local system
 $H^{(2)}$ is related with the observer's time $t_O$ as in (Time 2)
 of subsection I.3.2:
$$
t = t_O\sqrt{1-(2gL)/c^2} \approx t_O(1-(gL)/c^2).
$$
Therefore
$$
\exp(-itH^{(2)})f_2\approx \exp(-it_O\cdot H^{(2)})
\exp(it_O\cdot (gL/c^2)H^{(2)})f_2.
$$
We note that we can regard $H^{(1)}=H^{(2)}$ by
 $H=H^{(1)}+T^{(1)}=H^{(2)}+T^{(2)}$ and
$T^{(1)}=T^{(2)}=T$. As in (Energy 2) of subsection I.3.2,
 the internal energy $H^{(1)}=H^{(2)}$ of the system N+C
 or N+B is then approximated by a classical quantity
 $\mu c^2=m\left(1-\frac{m}{m+M}\right)\approx mc^2$.

For the first factor $\exp(-itT)$ on the right hand side of
 (E1), the time $t$ in the exponent is the local time of the
 local system $L$ as in (Time 1) of subsection I.3.2, because
 $T=T^{(j)}$ is the total intercluster free energy $T_{b^{(j)}}$
 corresponding to the cluster decomposition $b^{(j)}$ of the
 local system $L$. Since the center of mass of the local system
 $L$ is at the middle height between B and C, the time $t$ in
 $\exp(-itT)$ is thus related with $t_O$ as follows:
$$
t=t_O\sqrt{1-(gL)/c^2}.
$$

  From these, we have the following decomposition of the observed
 wave function for this 3-body system:
$$
\exp(-itH)f\approx \exp(-it_O\sqrt{1-(gL)/c^2}T)
\exp(-it_Omc^2)\{f_1+\exp(it_O\cdot gLm)f_2\}.
$$
Setting
$$
h_k(t_O)=\exp(-it_O\sqrt{1-(gL)/c^2}T)\exp(-it_Omc^2)f_k,\quad (k=1,2)
$$
we then have
$$
\exp(-itH)f\approx h_1(t_O)+\exp(it_O\cdot gLm)h_2(t_O).
$$
Therefore at the time $t_O$ of observation, there remains
 the desired phase difference, which explains the interference
 observed in [Co]. Note that $T=T^{(j)}=\frac{(D_y^{(j)})^2}{2\nu}$
 does not play any essential role in this argument, therefore we
 did not replace it by a classical quantity.

\BP

\subhead\nofrills{II.2. Hubble's law}
\endsubhead

\BP

\F
Hubble's law is a phenomenon that appears when one observes
 the light emitted from stars and galaxies far away from the earth.
 The emission of light itself is a quantum-mechanical phenomenon that
 could be explained by the nonrelativistic quantum field theory as
 in [Ki], Section 11-(2). The observation or reception of this
 emission of light on the earth is explained as a classical observation
 according to our postulate \linebreak Axiom 6.2, by introducing
 Robertson-Walker metric.

Robertson-Walker metric is the metric derived from the assumptions
 of {\it homogeneity} and {\it isotropy} of the large scale structure
 of the universe.
We refer the reader to [M], Chap. 27 for the details,
 and we here only outline the argument.

Under the hypotheses of homogeneity and isotropy, the metric is given
 in general as follows:
$$
ds^2=-(dx^0)^2+d\sigma^2=-(dx^0)^2+a(x^0)^2\gamma_{ij}(x^k)dx^idx^j,
$$
where $x^0$ is the time parameter that `slices' the spacetime by means
 of a one parameter family of some spacelike surfaces, and
 $(x^1,x^2,x^3)$ is the `comoving, synchronous space coordinate
 system' for the universe, in the sense of [M], sections
 27.3--27.4. $a(x^0)$ is the so-called ``expansion factor"
 that describes the ratio of expansion of the universe in the
 usual context of general theory of relativity.  A consideration by the
 use of homogeneity and isotropy
yields ([M], section 27.6) that for some functions $f(r)$
 ($r=|(x^1,x^2,x^3)|$) and $h(x^0)$
$$
ds^2=-(dx^0)^2+e^{f(r)}e^{h(x^0)}\{(dx^1)^2+(dx^2)^2+(dx^3)^2\}.
$$
Assuming Einstein field equation
$G^\mu_{\ \nu}-\lambda\delta^\mu_{\ \nu}=\kappa T^\mu_{\ \nu}$
 and calculating, we get with replacing $e^{h(x^0)}$ by a constant
 times $e^{h(x^0)}$
$$
ds^2=-(dx^0)^2+e^{h(x^0)}\left(1+k\frac{r^2}{4r_0^2}\right)^{-2}
\{(dx^1)^2+(dx^2)^2+(dx^3)^2\},
$$
where $k=-1$, $0$, or $+1$. This is called Robertson-Walker metric.
Using the polar coordinates
$(r, \theta, \varphi)$ and setting $t=x^0$ and
$$
\frac{r}{r_0}=u, \quad R(t)=r_0e^{h(t)/2},
$$
 one can rewrite $ds^2$ as follows:
$$
ds^2=-(dt)^2+R(t)^2\left(1+\frac{k}{4}u^2\right)^{-2}
[du^2+u^2\{(d\theta)^2+(\sin\theta d\varphi)^2\}].
$$

Suppose $k=+1$, and consider a 3-dimensional sphere of radius
 $A$ in a 4-dimensional Euclidean space
$$
A^2=(y^4)^2+\sum_{k=1}^3(y^k)^2.
$$
The metric on this sphere is
$$
d\sigma^2=\sum_{k=1}^3(dy^k)^2+(dy^4)^2.
$$
This is rewritten by using the equation of the sphere in the
 above as follows:
$$
d\sigma^2=\sum_{k=1}^3(dy^k)^2+
\left\{A^2-\sum_{k=1}^3(y^k)^2\right\}^{-1}
\left(\sum_{\ell=1}^3y^\ell dy^\ell\right)^2.
$$
Set $\rho^2=\sum_{k=1}^3(y^k)^2$, and define $v$ by
$$
\rho=A\left(1+\frac{v^2}{4}\right)^{-1}v.
$$
Using polar coordinates
$(\rho,\theta,\varphi)$
instead of $(y^1,y^2,y^3)$,
and rewriting $\rho$ by the use of $v$, we have
$$
d\sigma^2=A^2\left(1+\frac{v^2}{4}\right)^{-2}
[(dv)^2+v^2\{(d\theta)^2+(\sin\theta d\varphi)^2\}].
$$
If we set $A=R(t)$, and identify $v$ as $u$,
this formula coincides with the space part $d\sigma^2$ of
 the aforementioned Robertson-Walker metric $ds^2$.

In this sense, the space part slice $t=$ constant of the spacetime
 can be regarded as a 3-dimensional sphere of radius $R(t)$
 in a 4-dimensional Euclidean space, hence $R(t)=r_0e^{h(t)/2}$
 can be regarded as the radius of the universe and may expand as
 $t$ grows. The cosmological redshift observed by Hubble [Hu] gives
 in this context that $R(t)$ is growing at present (see section 29.2
 of [M]), and this is interpreted as a proof of `expansion' of the
 universe. However, as we have seen, the `expansion' is a consequence
 of the identification of $R(t)$ in the Robertson-Walker metric $ds^2$
 with the radius of a sphere in a virtual 4-dimensional Euclidean
 space. In this sense, the growth of $R(t)$ in $ds^2$ does not imply
 the expansion of the universe in any other senses than it is an
 `interpretation.'

The `expansion' of this type does not contradict the stationary
 universe $\phi$ in quantum-mechanical sense specified in Axiom 1.
The `expansion' is an interpretation of the observation {\it with
 one observer's coordinate system fixed}. The quantum-mechanical
 stationary universe $\phi$ is the inner structure of its own and
 is independent of the observer's coordinate system. In this
 sense, the `expansion' is an `appearance,' which the universe
 takes under the `interference' of the observer to try to reveal
 its morphology. More philosophically stating, the past and the
 future do not exist unless one fixes a time coordinate. The
 `Big Bang' is an imagination under the {\it presumption} that
 the time coordinate exists {\it a priori}.
 Unless it is observed with assuming the existence of a
 time coordinate, the universe can be a stationary state.

Our theory is a reflection and a clarification of this
 supposition of the existence of time coordinate, adopted
 {\it implicitly} in almost all physical theories today.

Example of the previous section is an experiment of human
 size, and the one in this section is an observation of
 cosmological size. These two examples together with the one
 in subsection I.3.2 would indicate a unified treatment of physical
 phenomena from the microscopic size to the cosmological size.

\BP

\vskip18pt

\centerline{\bf Part III. Open Problems}

\BP

\vskip12pt

\F
 In this Part III we state some open problems related
 with our formulation of physics. Some of them
 are known problems, but do not seem to have been
 given solutions. We conclude with stating
 a final goal of our formulation.

\BP

\subhead\nofrills{III.1. Stability of Matter}
\endsubhead

\BP

\F
As we have seen in Part I, the relativistic Hamiltonian
 considered by E. H. Lieb and others (see [Li] and the
 references therein) has reasonable grounds under the
 assumption that gravitation is weak. It has been thought
 that the non-invariance with respect to Lorentz transformation
 is its fault. However, according to our formulation,
 the non-invariance is not a fault but has
 natural foundations as a Hamiltonian which describes
 observational facts.  

It is therefore meaningful to research the related spectral and
 scattering theory for the relativistic Hamiltonian
$$
\tH_L=\sum_{j=1}^k\left(c\sqrt{D_j^2+m_j^2c^2}-m_jc^2\right)+I_b(x_b,0)
-G\sum_{1\le i<j\le k}m_im_j/r_{ij},
\tag QMG
$$
which includes the electric potentials and gravitational
 potentials simultaneously.

The first problem to be treated in this field is the problem of
 the stability of matter. It is known certain facts about this
 problem unless the electric potentials and gravitational ones
 are present simultaneously: If gravitation is absent and $I_b$
 is of the form
$$
I_b = -e^2\sum_{j=1}^k z|x_j-R|^{-1}
+e^2\sum_{1\le i<j\le k}|x_i-x_j|^{-1},
$$
where $R$ is the position of the nucleus with $z$ number of protons,
it is known that the stability of the first kind is equivalent to
 the stability of the second kind, and that atoms are stable when
 $z\le 87$. If electric potentials are absent, it is shown ([Li-Ya])
 that Thomas-Fermi theory is asymptotically exact for fermions.
 However, nothing seems known for the case which includes both of
 the electric potentials and gravitational ones as Lieb [Li] writes.
 The research to include both of electric and gravitational potentials
 would lead us to a deeper understanding of the nature of matter,
 since any matter includes both kinds of internal forces. E.g.,
 the stability and instability of stars with large number of
 particles would be understood in a more satisfactory manner
 than in the present understanding.

\BP

\vskip12pt


\subhead\nofrills{III.2. Scattering Theory}
\endsubhead

\BP

\F
The second problem to be considered would be the scattering theory
 for the Hamiltonian in the formula (QMG). This Hamiltonian has the
 potentials which are of Coulomb type, therefore, of critical
 singularity with respect to the free part:
$$
\tH_{L0}=\sum_{j=1}^k\left(c\sqrt{D_j^2+m_j^2c^2}-m_jc^2\right).
$$
Hence the problem of self-adjointness arises in the first place,
 and this is closely related with the problem of stability
 proposed in the previous section. The point is to what extent
 the free part $\tH_{L0}$ and the positive part of the sum of
 the potentials suppress the bad behavior of the negative parts.
 The gravitational potentials are quite small compared to
 the electric part and is negligible in the usual human size.
 But in the size of stars they cannot be neglected, and we
 have to develop some method which is able to treat the electric
 and gravitational parts at a time. This is the first problem
 which we should research in the scattering theory for (QMG).

If some conditions are established for the self-adjointness of
 (QMG), we should go on to the scattering phenomena governed by
 the Hamiltonian (QMG). This would give us an image about the
 phenomena which would be observed when the gravity and
 electrical forces are present simultaneously.

 At first glance, it looks as if there were a problem
 in our formulation in the point that the approximate
 relativistic Hamiltonian (QMG) would lose the self-adjointness
 and stability when the number of particles becomes large.
 This should, however, be taken as an evidence of the success
 of our formulation to include gravity. The fact that the universe
 does not seem to be subject to Boltzmann's heat death, but it,
 which is in the usual physical context supposed to have been
 in an equilibrium originally, could develop hot stars, is
 owing to the instability of gravitation. Our Hamiltonian (QMG)
 explains this fact so that our inclusion of gravity into (QMG)
 would be a reasonable one to that extent.

The final problem in this direction is to find a full general
 relativistic Hamiltonian, which explains the observation
 without assuming that the gravitation is weak. This seems
 difficult seeing the present stage of the theory, but we hope
 that this end would be accomplished in the future.

\BP

\BP


\vskip10pt

\Refs
\widestnumber\key{888888}

\ref \key A-M
\by R. Abraham and J. E. Marsden \book Foundations of Mechanics
\publ The Benjamin/Cummings  Publishing  Company, 2nd ed.
\publaddr London-Amsterdam-Don Mills, Ontario-Sydney-Tokyo
\yr 1978
\endref

\ref \key Bo
\by M. Born \paper Zur Quantenmechanik der Stossvorg\"ange
\jour Zeitshrift f\"ur Physik \yr 1926 \pages 863-867 \vol 37
\endref

\ref \key Co
\by R. Collela, A. W. Overhauser and S. A. Werner
\paper Observation of gravitationally induced quantum mechanics
\jour Phys. Rev. Lett. \vol 34 \yr 1975 \pages 1472-1474
\endref

\ref \key Cy
\by H. L. Cycon {\it et al.}
\book Schr\"odinger Operators
\publ Springer-Verlag
\yr 1987
\endref

\ref \key De
\by J. Derezi\'nski \paper Asymptotic completeness of long-range
$N$-body quantum systems
\yr 1993 \vol 138 \jour Annals of Math. \pages 427-476
\endref

\ref \key Di
\by P. A. M. Dirac \jour Proc. Roy. Soc. \vol A117, \pages 610
\yr 1928
\endref

\ref \key Dy
\by F. J. Dyson \yr 1953
\paper Divergence of perturbation theory in quantum electrodynamics
\jour Phys. Rev. \vol 75 \pages 486
\endref

\ref \key Ein
\by A. Einstein \yr 1923
\paper The foundation of the general theory of relativity
\inbook Translated by W. Perrett and G. B. Jeffery,
The Principle of Relativity \pages 111-164
\publ Dover
\endref

\ref
\key En \by V. Enss
\paper Introduction to asymptotic observables for multiparticle
 quantum scattering \inbook Schr\"od-inger Operators,
 Aarhus 1985, Ed. by E. Balslev, Lect.  Note in  Math.
 \vol 1218 \publ Springer-Verlag
\yr 1986 \pages 61-92
\endref

\ref \key Fr
\by J. Fr\"ohlich \yr 1982 \paper On the triviality of 
$\lambda\Phi_d^4$ theories and the approach to the critical point in
 $d\le 4$ dimensions \jour Nucl. Phys. \vol B 200 [FS4] 
\pages 281-296
\endref

\ref \key Hu
\by E. P. Hubble \paper A relation between distance and radial
 velocity among extragalactic nebulae
\jour Proc. Nat. Acad. Sci. U.S. \vol 15 \yr 1929 \pages 169-173
\endref

\ref \key Ish
\by C. J. Isham \yr 1993  \paper Canonical quantum gravity and the
 problem of time \inbook  Proceedings of the NATO Advanced Study
 Institute, Salamanca, June 1992 \publ Kluwer Academic Publishers
\endref

\ref \key I-Ki
\by H. Isozaki and H. Kitada \yr 1986 \paper Scattering matrices
 for two-body Schr\"odinger operators \jour Scientific Papers of
 the College of Arts and Sciences, The University of Tokyo 
\vol 35 \pages 81-107
\endref

\ref \key K-L
\by T. Kinoshita and W. B. Lindquist \yr 1983
\paper Eighth-order magnetic moment of the electron
\jour Phys. Rev. \vol D27
\pages 866
\endref

\ref \key Ki
\by H. Kitada \yr 1994  \paper Theory of local times 
\jour Il Nuovo Cimento
\vol 109 B, N. 3 \pages 281-302
\endref



\ref \key Ki($N$)
\by H. Kitada \paper Asymptotic completeness of N-body wave
operators  II. A new proof for the short-range case and the
 asymptotic clustering for long-range systems 
\inbook Functional Analysis and Related Topics, 1991,
 Ed. by H. Komatsu, Lect. Note in Math. \vol 1540
 \publ Springer-Verlag \yr 1993 \pages
 149-189
\endref

\ref \key Ki-Fl
\by H. Kitada and L. Fletcher \paper Local time and the unification
 of physics, Part I: Local time \jour Apeiron
\vol 3 \pages 38-45 \yr 1996
\endref


\ref \key Li
\by E. H. Lieb \paper The stability of matter: From atoms to stars
\jour Bull. Amer. Math. Soc. \vol 22 \pages 1-49 \yr 1990
\endref

\ref \key Li-Ya
\by E. H. Lieb and H-T. Yau \paper The Chandrasekhar theory of
 stellar collapse as the limit of quantum mechanics
\jour Commun. Math. Phys. \vol 112 \pages 147-174 \yr 1987
\endref


\ref \key M
\by C. W. Misner, K. S. Thorne, and J. A. Wheeler \book Gravitation
\publ W. H. Freeman and Company, New York \yr 1973
\endref

\ref \key New
\by I. Newton  \paper Sir Isaac Newton Principia, Vol. I The Motion
 of Bodies, Motte's translation Revised by Cajori \inbook Tr. Andrew
 Motte ed. Florian Cajori \publ Univ. of California Press, Berkeley,
 Los Angeles, London \yr 1962
\endref

\ref \key Pru
\by E. Prugove\v cki
\book Quantum Geometry, A Framework for Quantum General Relativity
\publ Kluwer Academic Publishers, Dordrecht-Boston-London
\yr 1992
\endref

\ref \key St
\by F. Streater \paper\nofrills Why should anyone want to axiomatize
 quantum field theory? 
\inbook Philosophical Foundations of Quantum Field Theory,
 Ed. by H. R. Brown and R. Harr\'e 
\publ Clarendon, Oxford University Press \pages 137-148 \yr 1990
\endref



\endRefs
\enddocument